\documentclass{jpp}
\usepackage{graphicx}
\usepackage{xcolor}
\usepackage{subcaption}
\usepackage[utf8]{inputenc}
\usepackage[T1]{fontenc}
\usepackage{amsmath}

\shorttitle{Energetic bounds on gyrokinetic instabilities. Part $4$}
\shortauthor{P. J. Costello and G. G. Plunk}

\title{Energetic bounds on gyrokinetic instabilities. Part 4. Bounce-averaged electrons.}

\author{P. J. Costello\aff{1}
  \corresp{\email{paul.costello@ipp.mpg.de}},
  G. G. Plunk\aff{1}}

\affiliation{\aff{1}Max-Planck-Institut f\"ur Plasmaphysik, Wendelsteinstraße 1, 17491 Greifswald, Germany}

\begin{document}

\maketitle

\begin{abstract}
Upper bounds on the growth of instabilities in gyrokinetic systems have recently been derived by considering the optimal perturbations that maximise the growth of a chosen energy norm. This technique has previously been applied to two-species gyrokinetic systems with fully kinetic ions and electrons. However, in tokamaks and stellarators, the expectation from linear instability analyses is that the most important kinetic-electron contribution to ion-scale modes comes from the trapped electrons, which bounce faster than the timescale upon which instabilities evolve. As a result, a fully-kinetic electron response is not required to describe unstable modes in most cases. Here, we apply the optimal mode analysis to a reduced two-species system that consists of fully gyrokinetic ions and bounce-averaged electrons with the aim of  finding a tighter bound on ion-scale instabilities in toroidal geometry. This analysis yields bounds that are greatly reduced in comparison to the earlier two-species result. Moreover, if the energy norm is properly chosen, wave-particle resonance effects can be captured, reproducing the stabilisation of density-gradient-driven instabilities in maximum-$J$ devices. The optimal mode analysis also reveals that the maximum-$J$ property has an additional stabilising effect on ion-temperature-gradient-driven instabilities, even in the absence of an electron-free energy source. This effect is explained in terms of the concept of mode inertia, making it distinct from other mechanisms.
\end{abstract}

\section{Introduction}
In a recent series of papers \citep{Helander-2021,helanderEnergeticBoundsGyrokinetic2022, plunkEnergeticBoundsGyrokinetic2022,plunkEnergeticBoundsGyrokinetic2023a} a new theory has been developed that bounds the growth of linear and nonlinear instabilities in gyrokinetic systems by considering the energy balance that must be obeyed by these systems (from now on we will refer to these publications as Parts I, II and III, respectively). These upper bounds are attained by the \textit{optimal modes} of the system -- states of the gyrokinetic system that maximise the growth of a chosen `energy norm' within the limits allowed by energy balance. The optimal modes form a complete orthogonal basis for the space of distribution functions. Unlike the `normal' modes of the linear gyrokinetic operator (solutions to the gyrokinetic system which evolve as $\exp({-i \omega t})$), the velocity space dependence of the optimal modes is significantly simpler. Moreover, the optimal modes possess a closed-form solution in terms of a finite set of gyro-fluid-like moments, without any closure approximation, resulting in a low-dimensionality eigenvalue problem for the optimal growth rate, which can be solved at low computational cost. This makes the optimal modes an attractive lens with which to explore gyrokinetic systems.

There is a wide range of energy norms that can be used in these optimal mode analyses. In Part I and Part II the bounds were developed using the Helmholtz free energy as an energy norm, culminating in upper bounds on fully-electromagnetic, multi-species instabilities. Part III extended the optimal mode analysis by considering the generalised free energy, a linear combination of the Helmholtz free energy and the electrostatic energy, for a single kinetic species (kinetic ions with adiabatic electrons). The optimal growth of this generalised free energy yields the lowest possible upper bound that can be formed from the balance of these two energies, as will become clearer below. The generalised free energy introduced in Part III adds an explicit dependence on magnetic geometry into the optimal mode analysis by retaining the wave-particle interaction effects, including magnetic drifts, that contribute in a physically transparent way to the evolution of the field energy of the system.

While the bounds derived for the system with kinetic ions and adiabatic electrons have been found to be quite tight  to the results of  linear gyrokinetic simulations in tokamaks and stellarators, the upper bounds for two kinetic species in the electrostatic limit derived in Part II tends to be several times larger than the simulated linear instability growth in these geometries \citep{podaviniEnergeticBoundsGyrokinetic2024}. 
This is most evident at small values of the perpendicular wave number $k_\perp$, where the simulated growth rates in the toroidal geometries tend to zero, and the upper bound tends to a non-zero value of considerable size. 
This discrepancy in qualitative behaviour is because the upper bound of Part II is geometry independent and must also bound the growth of instabilities  in closed-field-line geometries, like the z-pinch \citep{ricciGyrokineticLinearTheory2006}, which exhibit finite growth as $k_\perp \to 0$. As we will see, the difference in these geometries lies in the transit average of the passing-electron response, which may be non-adiabatic in closed-field-line geometry but is largely adiabatic on the typical instability timescale in toroidal geometry, where the trapped electrons dominate the non-adiabatic electron response.

In this work, we aim to increase the applicability of the optimal mode theory with a kinetic electron response to stellarator and tokamak geometry. This is done by considering a reduced two-species system, in the electrostatic limit, in which the electron transit time along the magnetic field is assumed to be much faster than the evolution of the instabilities. In toroidal geometry, this eliminates the passing-electron contribution to instabilities, at lowest order, leaving only a bounce-averaged trapped electron response. The resulting gyrokinetic system is often used to describe electrostatic instabilities such as trapped-electron modes (TEMs) \citep{helanderCollisionlessMicroinstabilitiesStellarators2013}, ion-driven TEMs (ITEMs) \citep{plunkCollisionlessMicroinstabilitiesStellarators2017}, but is also applicable to ion-temperature gradient modes (ITGs) \citep{prollTurbulenceMitigationMaximumJ2022}.  We note that we exclude modes such as the electron-temperature-gradient mode (ETG)  \citep{dorlandElectronTemperatureGradient2000, plunkStellaratorsResistTurbulent2019} and the universal mode \citep{landremanGeneralizedUniversalInstability2015, helanderUniversalInstabilityGeneral2015} which evolve on timescales comparable to, or shorter than, the electron transit time.

To include wave-particle resonance effects, particularly involving the magnetic drifts, we construct the generalised free energy (of which the Helmholtz free energy is a special case) of the system with gyrokinetic ions and bounce-averaged electrons. We first study the Helmholtz free energy limit, which is directly comparable with the two-species result of Part II.  We then consider the full generalised free energy and explore the impact of magnetic curvature on the optimal modes and the resulting upper bounds. We confirm that the previously known linear stability benefits of maximum-$J$ magnetic geometries associated with the presence of an electron free-energy source \citep{helanderCollisionlessMicroinstabilitiesStellarators2013, prollResilienceQuasiIsodynamicStellarators2012}  is exhibited by the optimal modes. The optimal mode analysis also reveals an additional stabilising effect that the maximum-$J$ property has on ITGs -- even in the absence of an electron-free-energy source. We explain this effect through the concept of mode inertia, which is sensitive to the degree of depletion of the Boltzmann response by the trapped electrons. 
This is distinct from other mechanisms described in previous works that enhance microstability in maximum-$J$ stellarators, which involve a lack of resonance between bounce-averaged curvature and diamagnetic drifts.

\section{Gyrokinetic system}\label{sec:equations}
To begin, we consider a local, electrostatic gyrokinetic system, where the domain is a narrow flux tube centred around a magnetic field line in the plasma. In this local approximation, a Fourier decomposition may be applied in the directions perpendicular to the magnetic field, assuming periodicity in these coordinates. The electrostatic gyrokinetic equation for a species $a$, and the Fourier component $\mathbf{k}$ is
\begin{multline}
    \frac{\partial g_{a, \mathbf{k}}}{\partial t} + v_{\|}\frac{\partial g_{a, \mathbf{k}}}{\partial l} + i\omega_{da} g_{a, \mathbf{k}}+ \frac{1}{B^2}\sum_{\mathbf{k}'}\mathbf{B} \cdot (\mathbf{k}\times \mathbf{k}')\delta\phi_\mathbf{k'}J_{0a} \ g_{a,\mathbf{k}-\mathbf{k}'} \\
    = \frac{e_a F_{a0}}{T_a}\left(\frac{\partial}{\partial t} + i\omega_{*a}^T\right)\delta \phi_\mathbf{k} J_{0a},
    \label{eqn:GK_equation}
\end{multline}
where $g_{a,\mathbf{k}}$ is the non-adiabatic part of the perturbed distribution function $g_a(\mathbf{R}, E_a, \mu_a, t) = \delta f(\mathbf{r}, \mathbf{v}, t) + e_a\delta \phi(\mathbf{r}, t) F_{a0}/T_a$,  $F_{a0}$  is the background Maxwellian distribution and $\delta \phi$ is the electrostatic potential. The phase-space coordinates for the gyro-centre distribution function $g_a$ are the magnetic moment, $\mu_a = m_a v_\perp^2/(2B)$ and the kinetic energy, $E_a = m_a v^2/2$.  We  also use Clebsch magnetic coordinates, in which the magnetic field is of the form $\mathbf{B} = \nabla \psi \times \nabla \alpha$, where $\psi$ is the toroidal magnetic flux (acting as a radial coordinate in toroidal geometry) and $\alpha$ is the binormal coordinate, which labels different field lines on surfaces of constant $\psi$. In these coordinates, the field-line following coordinate is  $l$, the unit vector along the magnetic field is $\mathbf{b} = \mathbf{B}/B$ and the perpendicular wavenumber is $\mathbf{k} = k_\alpha \nabla \alpha + k_\psi\nabla \psi$. The argument of the Bessel function is $J_0 = J_0(k_\perp v_\perp/\Omega_a)$ with the gyrofrequency defined as $\Omega_a = e_a B/m_a$.
The magnetic drift frequency is $\omega_{da}= \mathbf{k}_\perp\cdot \mathbf{v}_d$, where $\mathbf{v}_d = {\mathbf{b}}\times(v_\perp^2/2 \nabla \ln B + v_\|^2 \mathbf{\kappa})/{\Omega_a}$. In the low-plasma pressure limit (assuming $\nabla \ln B \approx \mathbf{\kappa}$), the drift frequency can be approximated by
\begin{equation}
    \omega_{da} \approx \hat{\omega}_{da}(l)\left( \frac{v_\perp^2}{2 v_{Ta}^2} + \frac{v_\|^2}{v_{Ta}^2}\right),
\end{equation}
where $v_{Ta} = \sqrt{2T_a/m_a}$ is the thermal speed and $\tilde{\omega}_{da}(l)$ is a geometry dependent factor.  The energy-dependent diamagnetic drift frequency is
\begin{equation}
    \omega^T_a = \omega_{*a}\left[1 + \eta_a\left(\frac{v^2}{v_{Ta}^2} - \frac{3}{2}\right)\right],
\end{equation}
where the plasma gradients enter via $\omega_{*a} = (k_\alpha T_a/e_a)\mathrm{d} \ln n_a /\mathrm{d} \psi$ and the gradient ratio $\eta_a = \, \mathrm{d}\ln T_a/\, \mathrm{d} \ln n_a$. 
In the electrostatic limit, the system is closed by the quasi-neutrality condition
\begin{equation}
    \sum_a \frac{e_a^2 n_a}{T_a}\delta \phi_\mathbf{k} =  \sum_a e_a\int g_{a,\mathbf{k}} J_{0a} \, \mathrm{d}^3v.
\end{equation}
In Clebsch coordinates, we may write the nonlinear term of  \eqref{eqn:GK_equation} in the more convenient form
\begin{equation}
\frac{1}{B^2}\sum_{\mathbf{k}'}\mathbf{B} \cdot (\mathbf{k}\times \mathbf{k}')\delta\phi_\mathbf{k'}J_{0a} \ g_{a,\mathbf{k}-\mathbf{k}'}  =   \sum_{\mathbf{k}'} \left(k_\psi k_\alpha' -k_\alpha k_\psi' \right)\delta\phi_\mathbf{k'}J_{0a} \ g_{a,\mathbf{k}-\mathbf{k}'}.  
\end{equation}

\section{Bounce-averaged electrons}
To simplify the electron gyrokinetic equation, we employ a standard ordering.
We assume that the timescale of  the dynamics of interest in the gyrokinetic system $\tau_D$ (i.e the growth time of an instability) is much longer than the timescale on which thermal electrons transit the typical parallel length scale of the system, such that $\tau_D \gg L/v_{Te}$.  Expanding the electron distribution in orders of $L/(v_{Te}\tau_D)$ as $g_{e,\mathbf{k}} = g_{e,\mathbf{k}}^0 + g_{e,\mathbf{k}}^1 \ldots$ we find that, to leading order,
\begin{equation}
    v_\parallel \frac{\partial g_{e,\mathbf{k}}^0}{\partial l} = 0,
\end{equation}
such that the lowest-order electron distribution function is \textit{constant along the magnetic field line}. This, it turns out, is a highly consequential result of this ordering, and how it is interpreted is the key to tailoring the forthcoming optimal mode analysis to capture the behaviour of the electron response for ion-scale instabilities that satisfy our ordering in toroidal geometry. 

The constancy of  $g_e = g_e(E_e,\mu_e, \psi, \alpha)$ in $l$ implies that its value is entirely determined by the parallel boundary conditions. For the passing particles in toroidal geometry, and $k_\alpha \neq 0 $, the relevant boundary conditions are the so-called `incoming' boundary conditions of ballooning space where,
\begin{eqnarray}
     g_{e,\mathbf{k}}(v_\parallel > 0, l = -\infty) &=& 0, \\
          g_{e,\mathbf{k}}(v_\parallel < 0, l = \infty) &=& 0.
\end{eqnarray}
As explained in an appendix by \citet{plunkCollisionlessMicroinstabilitiesStellarators2014}, these boundary conditions are required to preserve causality \citep{connorStabilityGeneralPlasma1980}, such that an information from the boundary, which is infinitely far way, cannot be communicated in a finite time. Therefore, it must be the case that $g_{e}(k_\alpha \neq 0) = 0$ in the passing region of velocity space to satisfy the boundary conditions. For the $k_\alpha = 0$ component, ballooning space boundary conditions are not appropriate because there is no mechanism that should cause the mode to decay along the flux tube. Thus, for $k_\alpha = 0$ the electron distribution function has a zeroth order contribution from the passing electrons.

The trapped particles on the other hand (in the absence of collisions), are confined to a limited region of ballooning space. As such, the appropriate boundary conditions for these particles are the `turning-point' boundary conditions \citep{connorStabilityGeneralPlasma1980},
\begin{eqnarray}
     g_{e,\mathbf{k}}(v_\parallel > 0, l = l_1) &=& g_{e,\mathbf{k}}(v_\parallel < 0, l = l_1), \\
          g_{e,\mathbf{k}}(v_\parallel > 0, l = l_2) &=& g_{e,\mathbf{k}}(v_\parallel < 0, l = l_2).
\end{eqnarray}
These boundary conditions can be satisfied with $g_{e,\mathbf{k}} \neq 0$, and so, for $k_\alpha \neq 0$, only the trapped electrons contribute to the non-adiabtic electron response in this limit. Note that this is not the case in closed-field-line geometries where the parallel boundary conditions can be periodic, allowing $g_{e,\mathbf{k}}^0$ to be non-zero in the passing region of velocity space to zeroth order. 

At next order in $L/(v_{Te}\tau_D)$, we find,
\begin{multline}
\label{eqn:1st_order_electron_gk}
    \frac{\partial g_{e,\mathbf{k}}^0}{\partial t} + v_\parallel \frac{\partial g_{e,\mathbf{k}}^1}{\partial l} + i \omega_{de}g_{e,\mathbf{k}}^0 + \sum_{\mathbf{k}'} \left(k_\psi k_\alpha' -k_\alpha k_\psi' \right)\delta\phi_{\mathbf{k}'}J_{0e}  g_{e,\mathbf{k}-\mathbf{k}'}^0 =
    \\
    -\frac{e F_{e0}}{T_e}\left(\frac{\partial}{\partial t} + i\omega_{*e}^T\right)\delta \phi_\mathbf{k} J_{0e}.   
\end{multline}
We now define the bounce average as
\begin{equation}
    \overline{(\ldots)} = \frac{\int_{l_1}^{l_2} (\ldots) {\, \mathrm{d}l}/{v_\|}}{\int_{l_1}^{l_2}{\, \mathrm{d}l}/{v_\|}} = \frac{1}{\tau_B}\int_{l_1}^{l_2} \frac{\, \mathrm{d}l}{\sqrt{1 -\lambda B}},
\end{equation}
where we have introduced the pitch angle $\lambda = v_\perp^2/(v^2 B)$ and $\tau_B = \int_{l_1}^{l_2}\frac{\, \mathrm{d}l}{\sqrt{1 -\lambda B}}$, the bounce time\footnote{  $\tau_B$ here has units of length, due to the cancellation of $v$ when changing to pitch-angle coordinates, but is typically referred to as the bounce time for trapped particles.}. In pitch-angle coordinates, the velocity space element is defined as,
\begin{equation}
\label{eqn:Pitch_angle_element}
    \, \mathrm{d}^3 v = \sum_\sigma \frac{\pi v^2 B \, \mathrm{d} v \, \mathrm{d}\lambda}{\sqrt{1 -\lambda B}},
\end{equation}
where $\sigma = v_\parallel/|v_\parallel|$. Applying the bounce average to \eqref{eqn:1st_order_electron_gk}  annihilates the $g_{e, \mathbf{k}}^1$
contribution, leaving an equation for $g^0_{e,\mathbf{k}}$. Additionally ordering $k_\perp\rho_e\ll 1$, where $\rho_a = v_{Ta}/|\Omega_a|$, for simplicity and omitting the superscript `$0$' for brevity gives,
\begin{equation}
\label{eqn:Bounce_average_electron_gk}
    \frac{\partial g_{e,\mathbf{k}}}{\partial t} + i \overline{\omega}_{de}g_{e,\mathbf{k}}+ \sum_{\mathbf{k}'} \left(k_\psi k_\alpha' -k_\alpha k_\psi' \right)\overline{\delta \phi}_\mathbf{k'} \ g_{e,\mathbf{k}-\mathbf{k}'} =
    -\frac{e F_{e0}}{T_e}\left(\frac{\partial}{\partial t} + i\omega_{*e}^T\right)\overline{\delta \phi}_\mathbf{k}.   
\end{equation}
We now consider \eqref{eqn:Bounce_average_electron_gk} as the governing equation for the bounce-averaged electrons in our gyrokinetic system, which is accurate to lowest order in $L/(v_{Te}\tau_D)$.  Here, we restrict our view to toroidal geometry such that, for  $k_\alpha \neq 0$, $g_e$ is only non-zero in the trapped region of velocity space. 


\section{Energy Balance}
In the limit of  $\tau_D \gg L/v_{Te}$, we have a gyrokinetic system comprised of ions described by \eqref{eqn:GK_equation}, with finite ion-Larmor-radius effects, and bounce-averaged electrons described by \eqref{eqn:1st_order_electron_gk}.  As described in Parts I, II and III, we can construct energy norms for this two-species system which are conserved by the nonlinear terms of the gyrokinetic equations. These nonlinear invariants are the Helmholtz free energy, detailed in Parts I and II, and the electrostatic energy, described in Part III.
We now derive these invariants in our system with bounce-averaged electrons.
\subsection{Helmholtz free energy}
For simplicity, we will consider a two-species hydrogen plasma with $n_i = n_e = n$, but the following can be easily generalised to multiple ion species. Constructing the Helmholtz free energy of this system proceeds in the same manner as in the general multi-species case shown in Part I, except for our assumption that $k_\perp\rho_e\ll 1$. This is done by multiplying the gyrokinetic equation for each species by the factor $T_a g_{a,\mathbf{k}}^*/F_{a0}$, integrating over velocity space, applying the flux-tube average defined by
\begin{equation}
    \langle \ldots \rangle = \lim_{L\to \infty}{\int_{-L}^{L}(\ldots)\cfrac{\, \mathrm{d}l}{B}} \ \bigg/ {\int_{-L}^{L}\cfrac{\, \mathrm{d}l}{B}} =  \lim_{L\to \infty} \frac{1}{V} {\int_{-L}^{L}(\ldots)\cfrac{\, \mathrm{d}l}{B}},
\end{equation}
summing the equations for both species, and taking the real part. Applying the flux-tube average has no impact on the bounce-averaged electron equation, but we leverage the identity
\begin{equation}
    \bigg\langle \int \overline{f(\mathbf{v},l)} \ \, \mathrm{d}^3 v \bigg\rangle = \bigg\langle \int {f}(\mathbf{v},l)\  \, \mathrm{d}^3 v \bigg\rangle
\end{equation}
to allow us to combine the electron and ion contributions to the Helmholtz free energy.
After using quasi-neutrality, we arrive at the Helmholtz free energy, which looks similar to that of Parts I and II,
\begin{equation}
    {H}(\mathbf{k}, t) = \sum_{a} \bigg\langle T_a\int \frac{|g_{a,\mathbf{k}}|^2}{F_{a0}}\mathrm{d}^3 v - \frac{e_a^2 n}{T_a}| \delta \phi_\mathbf{k}|^2 \bigg\rangle. 
\end{equation}
The Helmholtz free-energy balance reads, as before,
\begin{multline}
    \sum_\mathbf{k} \frac{\, \mathrm{d}}{ \, \mathrm{d} t} H(\mathbf{k}, t) = 2 \operatorname{Re} \sum_{a, \mathbf{k}} \bigg\langle e_a \int i\omega_{*a}^T \delta \phi_\mathbf{k}  J_{0a}   g_{a,\mathbf{k}}^* \, \mathrm{d}^3v  \bigg\rangle \\ =  2 \sum_\mathbf{k} \left(D_i(\mathbf{k},t) + D_e^{\mathrm{tr}}(\mathbf{k},t)\right),
\end{multline}
the only difference here is that we have considered the limit $J_{0e} \approx 1$ and $g_e = g_e(E_e,\mu_e, \psi, \alpha)$ with $g_e = 0$ for $\lambda < 1/B_{\mathrm{max}}$. We have denoted the electron contribution with a `tr' superscript to signify that only the trapped region of velocity space contributes. The total free energy drive for a given $\mathbf{k}$ is then given by $D(\mathbf{k}, t) = D_i(\mathbf{k},t) + D_e^{\mathrm{tr}}(\mathbf{k},t)$, which determines the rate at which the system can grow given the presence of gradients in the plasma.

\subsection{Electrostatic energy}
To construct the generalised free energy, as demonstrated for a single kinetic species in Part III,  we first need to derive the electrostatic energy balance of the system with bounce-averaged electrons. As shown in Appendix A of Part III, the electrostatic energy of a multi-species system may be constructed by applying the operator
\begin{equation}
\label{eqn:electro_int_operator}
    \operatorname{Re}\sum_{\mathbf{k}} \bigg\langle \int   e_a \delta \phi_\mathbf{k}^* J_{0a}\ (\ldots) \mathrm{d^3}v\bigg\rangle
\end{equation}
to the gyrokinetic equations for each species and summing over species. When this operator is applied to the electron equation, the nonlinear term is annihilated. This can be shown by noticing that its contribution,
\begin{equation}
     \operatorname{Re}  \sum_{\mathbf{k},\mathbf{k}'} \bigg \langle \int \left(k_\psi k_\alpha' -k_\alpha k_\psi' \right)\delta \phi_{\mathbf{k}}^*\overline{\delta \phi}_{\mathbf{k}'} \ g_{e,\mathbf{k}-\mathbf{k}'} \, \mathrm{d}^3 v\bigg \rangle, 
\end{equation}
when expressed in pitch-angle coordinates \eqref{eqn:Pitch_angle_element} and using the identity \citep{helanderCollisionlessMicroinstabilitiesStellarators2013}
\begin{equation}
\label{eqn:dlambda_dl_identity}
    \int_{-\infty}^{+\infty} \, \mathrm{d} l \int_{1/B_{\mathrm{max}}}^{1/B(l)} \, \mathrm{d} \lambda \to \int_{1/B_{\mathrm{max}}}^{1/B_{\mathrm{min}}} \, \mathrm{d} \lambda \sum_j \int_{l_{1,j}}^{l_{2,j}} \, \mathrm{d}l ,
\end{equation}
where the subscript $j$ denotes the different magnetic trapping wells along the flux tube,  becomes,
\begin{equation}
      \operatorname{Re} \sum_{\mathbf{k},\mathbf{k}', \sigma, j} \frac{\pi}{V} \int_{1/B_{\mathrm{max}}}^{1/B_{\mathrm{min}}} \, \mathrm{d} \lambda \ \tau_{B,j} \int_{0}^{\infty} \, \mathrm{d} v \ v^2 \left(k_\psi k_\alpha' - k_\alpha k_\psi' \right)\overline{\delta\phi}_{\mathbf{k}}^*\overline{\delta \phi}_{\mathbf{k}'} \ g_{e,\mathbf{k}-\mathbf{k}'}.
\end{equation}
Since $\delta \phi^*_\mathbf{k} = \delta \phi_{-\mathbf{k}}$  and $g_{e,\mathbf{k}}^* = g_{e,-\mathbf{k}}$, this contribution is proportional to,
\begin{equation}
    \left(k_\psi k_\alpha' - k_\alpha k_\psi' \right)\left(\overline{\delta\phi}_{-\mathbf{k}}\overline{\delta \phi}_\mathbf{k'} \ g_{e,\mathbf{k}-\mathbf{k}'} + \overline{\delta\phi}_{-\mathbf{k}}\overline{\delta \phi}_\mathbf{k'} \ g_{e,-\mathbf{k}+\mathbf{k}'}\right)
\end{equation}
which changes sign if $\mathbf{k}$ and $\mathbf{k}'$ are swapped, such that this term vanishes upon summation over $\mathbf{k}$ and $\mathbf{k}'$. 

What remains of the electron equation when this operator is applied is,
\begin{multline}
       \operatorname{Re}\sum_{\mathbf{k}} \bigg\langle \int   -e \ \delta \phi_\mathbf{k}^*\left(\frac{\partial g_{e,\mathbf{k}}}{\partial t} + i \overline{\omega}_{de} g_{e,\mathbf{k}}\right) \mathrm{d^3}v\bigg\rangle \\ =   \operatorname{Re}\sum_{\mathbf{k}}\bigg\langle \frac{e^2}{T_e}\int F_{e0} \delta\phi^*_\mathbf{k}\left(\frac{\partial}{\partial t} + i \omega_{*e}^T \right)\overline{\delta \phi}_\mathbf{k} \mathrm{d^3}v \bigg\rangle. 
\end{multline}
We may also use the identity \eqref{eqn:dlambda_dl_identity} to rewrite the right-hand-side of  this expression. Upon taking the real part of the right-hand-side, the term proportional to  $\omega_{*e}^T$ vanishes leaving,
\begin{multline}
\label{eqn:electron_contribution_to_E}
       \operatorname{Re}\sum_{\mathbf{k}} \bigg\langle \int   -e  \delta \phi_\mathbf{k}^*\left(\frac{\partial g_{e,\mathbf{k}}}{\partial t} + i \overline{\omega}_{de} g_{e, \mathbf{k}}\right) \mathrm{d^3}v\bigg\rangle \\ =   \frac{1}{4 V}\frac{\, \mathrm{d}}{\, \mathrm{d} t} \sum_{\mathbf{k}} \frac{e^2 n}{T_e}\sum_j \int_{1/B_{\mathrm{min}}}^{1/B_{\mathrm{max}}} \tau_{B,j} |\overline{\delta \phi}_{\mathbf{k},j} |^2\, \mathrm{d} \lambda.
\end{multline}
Applying the operator \eqref{eqn:electro_int_operator} to the ion gyrokinetic equation proceeds similarly, albeit without bounce averages, yielding
\begin{multline}
\label{eqn:ion_contribution_to_E}
    \operatorname{Re}\sum_{\mathbf{k}} \bigg\langle \int   e  \delta \phi_\mathbf{k}^*J_{0i}\left(\frac{\partial g_{i,\mathbf{k}}}{\partial t}+ v_\parallel \frac{\partial g_i}{\partial l} + i {\omega}_{di} g_{i,\mathbf{k}}\right) \mathrm{d^3}v\bigg\rangle \\
    =   \frac{\, \mathrm{d}}{\, \mathrm{d} t } \sum_{\mathbf{k}}\frac{e^2 n}{T_i}\bigg\langle \Gamma_{0i}(b)|{\delta \phi}_\mathbf{k}|^2  \bigg\rangle,
\end{multline}
where $\Gamma_{0i}(b) = I_0(b)\exp(-b)$  and $b = k_\perp^2\rho_i^2$. If we now sum the contributions from the ions and the electrons, given by  \eqref{eqn:ion_contribution_to_E} and \eqref{eqn:electron_contribution_to_E}, after using quasi-neutrality, we arrive at electrostatic energy balance:
\begin{equation}
    \frac{\mathrm{d}}{\mathrm{d} t}\sum_\mathbf{k} E(\mathbf{k}, t) = 2 \sum_\mathbf{k}\left(K_i(\mathbf{k},t) + K_e^{\mathrm{tr}}(\mathbf{k},t) \right).
\end{equation}
Here, $E(\mathbf{k}, t)$ is electrostatic energy contained in the potential fluctuations in this system given by
\begin{equation}
    E(\mathbf{k},t) = \frac{e^2 n}{T_i}\left[\bigg\langle \left( 1 + \tau - \Gamma_{0i}(b)\right)|\delta \phi_\mathbf{k}|^2 \bigg\rangle- \frac{\tau}{2V}\sum_j \int_{1/B_{\mathrm{min}}}^{1/B_{\mathrm{max}}} \tau_{B,j} |\overline{\delta \phi}_{j,\mathbf{k}} |^2\, \mathrm{d} \lambda \right ],
\label{eqn:electrostatic_energy}
\end{equation}
where $\tau = T_i/T_e$. The drive terms $K_i(\mathbf{k}, t)$ and $K_e^{\mathrm{tr}}(\mathbf{k},t)$ are the work performed by the ions and trapped electrons on the electric field, respectively, and are given by
\begin{equation}
    K_i(\mathbf{k},t) = -\operatorname{Re} \bigg\langle e \int  \delta \phi_\mathbf{k}^*J_{0i}\left(v_\parallel \frac{\partial g_i}{\partial l} + i {\omega}_{di} g_{i,\mathbf{k}}\right) \mathrm{d^3}v\bigg\rangle,
\end{equation}
and
\begin{equation*}
    K_e^{\mathrm{tr}}(\mathbf{k},t) = \operatorname{Re} \bigg\langle e  \int i\overline{{\omega}}_{de}  \delta \phi_\mathbf{k}^* g_{e,\mathbf{k}} \mathrm{d^3}v\bigg\rangle.
\end{equation*}
The total electrostatic energy drive is then $K(\mathbf{k}, t) = K_i(\mathbf{k},t) + K_e^{\mathrm{tr}}(\mathbf{k},t)$. The drive, $K$, determines the rate at which the electrostatic energy of the system can grow, given the energy transfer possible by wave-particle interactions.
This form of electrostatic energy balance is similar to forms derived by \cite{helanderCollisionlessMicroinstabilitiesStellarators2013} and \cite{plunkCollisionlessMicroinstabilitiesStellarators2017}, where they were concerned with normal modes.

\subsection{Generalised free energy}
Now that we have found forms for both the electrostatic energy and the Helmholtz free energy, we can construct the generalised free energy, as shown in Part III, by considering the linear combination
\begin{equation}
    \tilde{H} = H - \Delta E,    
\end{equation}
where $\Delta$ is a real constant. $\tilde{H}$ can be guaranteed to be a positive-definite energy norm if $\Delta$ is chosen correctly; for example, any negative real choice of $\Delta$ gives a positive definite energy measure. However, positive-definiteness will not be fulfilled beyond a certain magnitude of  positive $\Delta$. The generalised free energy $\tilde{H}$ is also invariant under nonlinear interactions between different wave numbers if $\Delta$ is chosen to be independent of $k$  \citep{plunkEnergeticBoundsGyrokinetic2023a}.
If the value of $\Delta = 0$ is chosen, then the generalised free energy reduces to the Helmholtz free energy.
The energy balance equation for $\tilde{H}$ is
\begin{equation}
    \frac{\, \mathrm{d} }{\, \mathrm{d} t } \sum_\mathbf{k} \tilde{H}(\mathbf{k},t) = 2 \sum_\mathbf{k}\left(D(\mathbf{k},t)- \Delta K(\mathbf{k},t) \right). 
\end{equation}
This energy balance is now `aware' of both the \textit{source} of free energy growth (the gradients), and the \textit{mechanism} by which it is extracted (wave-particle interactions).
As defined in Part III, the instantaneous growth rate of generalised free energy  is,
\begin{equation}
\label{eqn:Lamda_definition}
    \Lambda  = \left(D - \Delta K \right)/\tilde{H}.
\end{equation}
\section{Optimal modes}
We now seek the distribution functions $g_i$ and $g_e$ which maximise the value of $\Lambda$ -- the optimal modes. These optimal modes are found by varying \eqref{eqn:Lamda_definition} with respect to the distribution function for each species and seeking an extremum ($\delta \Lambda = 0$). This optimisation procedure yields two coupled variational problems, one for each species, given by,
\begin{equation}
\label{eqn:variational_problem}
    \Lambda \frac{\delta \tilde{H}}{\delta g_a} = \frac{\delta D}{\delta g_a} -\Delta\frac{\delta K}{\delta g_a}.
\end{equation}
However, when performing the variation, we must take into consideration that the electron distribution function is independent of the field-line-following coordinate $l$. To enforce this, when computing the variation of a functional $F[g_e]$ as
\begin{equation}
    \frac{\delta  F}{\delta g_e} = \frac{d}{d\varepsilon}\bigg\rvert_{\varepsilon= 0}F[g_e + \varepsilon h_e],
\end{equation}
the perturbation $h_e$ is taken to be independent of $l$. To make the impact of $l$-independence clearer, we define the inner product on the space of trapped-electron distribution functions as
\begin{equation}
\label{eqn:electron_inner_prod}
 \left(g_e, g_e\right)_{tr} = \frac{2\pi}{V}\sum_{j}\int_{1/B_{\mathrm{max}}}^{1/B_{\mathrm{min}}} \, \mathrm{d} \lambda \int_{0}^{\infty} \, \mathrm{d} v  v^2 \tau_{B,j} T_e \frac{|g_e|^2}{F_{e0}},
\end{equation}
which can be derived from the inner product for ion distributions (which we also use for the ion distributions here) defined in Part III,
\begin{equation}
\label{eqn:ion_inner_prod}
    (g_i, g_i) = \bigg\langle T_i \int \  \frac{|g_i|^2}{F_{i0}} \, \mathrm{d}^3v  \bigg\rangle
\end{equation}
by expressing the velocity space integration in pitch-angle coordinates, under the assumptions of $l$-independence, and $g_e$ being zero in the passing region of velocity space. 

Evaluating the two variational problems \eqref{eqn:variational_problem}  and expressing the result in terms of the inner products defined above, for the respective species over which the variation is performed, yields two coupled eigenvalue problems for the optimal modes and the  growth rate $\Lambda$:
\begin{equation}
\label{eqn:kinetic_eigenvalue_problem}
    \Lambda \sum_b \tilde{\mathcal{H}}_{a b} g_b = \sum_b \left( \mathcal{D}_{a b}g_b - \Delta \mathcal{K}_{a b}g_b\right),
\end{equation}
where the operators $\tilde{\mathcal{H}}_{ab}$, ${\mathcal{D}}_{ab}$  and $\mathcal{K}_{ab}$ are given in Appendix~\ref{appendix:Hermitian_linear_ops}. Here, and from now on, we will omit the subscript `$\mathbf{k}$' for simplicity. These operators are expressed in terms of a finite set of velocity-space moments of the distributions:
\begin{eqnarray}
\label{eqn:moment_defs_1}
\kappa_{1a} &=& \frac{1}{n}\int g_a J_{0a} \, \mathrm{d}^3 v, \\
\kappa_{2a} &=& \frac{1}{n}\int \left( \frac{v^2}{v_{Ta}^2}\right) g_a J_{0a} \, \mathrm{d}^3 v,  \\
    \kappa_{3e} &=& \frac{1}{n}\int\overline{f(l)(1 - \lambda B /2)}\left( \frac{v^2}{v_{Te}^2}\right) g_e \, \mathrm{d}^3 v,  \\
    \kappa_{3i} &=& \frac{1}{n}\int\left( \frac{v_\perp^2}{2 v_{Ti}^2} + \frac{v_\parallel^2}{v_{Ti}^2}\right) g_i J_{0i} \, \mathrm{d}^3 v, \\
\kappa_{4i} &=& \frac{1}{n}\int \left( \frac{v_\parallel}{v_{Ti}}\right) g_i J_{0i} \, \mathrm{d}^3 v, \\
\kappa_{5i} &=& \frac{1}{n}\int \left( \frac{v_\parallel}{v_{Ti}}\right) g_i \frac{\partial J_{0i}}{\partial l} \, \mathrm{d}^3 v,
\label{eqn:moment_defs_2}
\end{eqnarray}
where we have written $\hat{\omega}_{de}(l) = \tilde{\omega}_{de} f(l)$, $f(l)$ is a dimensionless function which captures the $l$-dependence of $\hat{\omega}_{de}$, and $\tilde{\omega}_{de}$ is a constant with units of inverse time. Note that, due to the appearance of the bounce average in each of the electron operators in Appendix~\ref{appendix:Hermitian_linear_ops}, the equation which results from the variation of the electron distribution function gives an optimal $g_e$ which is independent of $l$ as required.

This eigenvalue problem for the optimal mode growth has `full' dependence on the magnetic geometry of the flux tube, i.e. the magnetic field strength, the curvature and the metric coefficients, all as a function of $l$. The system can be closed exactly in terms of the moments $\kappa_{n a}$by taking moments of the two equations given by \eqref{eqn:kinetic_eigenvalue_problem}. This results in an $8\times8$ system of integro-differential equations which must be solved for the eigenvalue $\Lambda$.   Note that, due to the complexity of the system in general magnetic geometry, we are unable to derive generally the limiting value of $\Delta$ beyond which positive definiteness is no longer guaranteed. However, this does not pose a problem when solving the system numerically as positive definiteness can be checked at each step when optimising over $\Delta$.

\section{Solving the system in the slow-ion-transit limit  }
To make finding the solution to the eigenvalue problem more straightforward we consider the limit $L/v_{Ti} \gg \tau_D$, where the ion transit time is much longer than the typical instability timescale. In this limit, we can neglect the derivatives along the magnetic field line which arise due to the parallel ion motion. This reduces the dimensionality of the system by removing $\kappa_{4i}$ and $\kappa_{5i}$ from the problem, leaving a $6 \times 6$ system of integro-algebraic equations. Note that this ordering is the typical `toroidal' ITG and TEM ordering often applied in linear, normal-mode analyses \citep{ biglariToroidalIonPressure1989, helanderCollisionlessMicroinstabilitiesStellarators2013, plunkCollisionlessMicroinstabilitiesStellarators2017}. The system of moment equations in this limit is given in Appendix \ref{appendix:Moment_forms}.
\subsection{A square magnetic well}
We cannot make much progress towards an analytical solution of this system for a general magnetic geometry. Thus, to gain some insight into the behaviour of the system, we consider a `square' magnetic field of the form
\[B(l) = \begin{cases} 
      B_{\mathrm{min}} & -L/2 < l < L/2 \\
      B_{\mathrm{max}} &  |l| \geq L/2 \\
   \end{cases}
\]
with $\omega_{da}(l) = \mathrm{const.}$ and $k_\perp = \mathrm{const}$. In this simplified geometry, the $\kappa_{na}$ eigenfunctions are constant inside the magnetic well between $-L/2 <l < L/2 $ and the $\kappa_{ne}$ moments must vanish for $l\geq L/2$. In this case, there are two possibilities, either the ion moments must also vanish in this region, or the electron moments are zero everywhere, corresponding to the system with adiabatic electrons. Both of these solutions are realisable, with the adiabatic electron solution being described in Part III. Here, we concern ourselves with the eigenfunctions that vanish at $B_\mathrm{max}$, whilst keeping in mind that the adiabatic electron solution is also present in the system outside the well.

The solution to the eigenvalue problem in this square magnetic geometry is given in Appendix \ref{appendix:square_well} and is denoted as $\Lambda_{\mathrm{SW}}$.  We find that the value of $\Delta$ beyond which the system is not guaranteed to be Hermitian (i.e implying a violation of positive-definiteness) is \begin{equation}
    \Delta < \frac{1+ \tau}{\Gamma_0 + \tau \varepsilon},
\end{equation}
where $\varepsilon = \sqrt{1 - B_{\textrm{min}}/B_{\textrm{max}}}$. In the limit of $B_\textrm{min}/B_\textrm{max} \to 1$, $\Lambda_{\textrm{SW}}$ reduces to the adiabatic electron result of Part III in the toroidal ITG limit.

\subsection{Numerically solving the system}
\label{sec:numerically_solving_the_system}
The analytical solution in the square well, $\Lambda_{\mathrm{SW}}$, already provides some insight into the behaviour of the eigenvalue problem but does not capture the subtleties of the bounce-averaged electron response. In realistic magnetic geometries, the local curvature along the field line may differ vastly from the bounce-averaged curvature felt by trapped electrons due to the variation of the magnetic field strength along the field line. As a result, the ions and electrons in the bounce-averaged limit can experience drifts which are vastly different. This could be captured in the square well geometry by  stipulating that the ions and electrons experience different curvatures in the well, but this would be a coarse approximation of the realistic situation.

To get a better sense of the behaviour of the system in more realistic geometries, we resort to numerically solving the system of integro-algebraic equations given by the moment form of the kinetic eigenvalue problem \eqref{eqn:kinetic_eigenvalue_problem} in the slow-ion-transit limit ($L/v_{Ti} \gg \tau_D$). For the sake of simplicity, we do this in simple, toy-model magnetic fields. Here, we discretise the system on a basis of cosines in $l$, with zeros at the maximum of magnetic field strength (the eigenfunctions of interest are purely even in a symmetric geometry, and must vanish at $B = B_{\mathrm{max}}$), and use a recently developed numerical package to evaluate the bounce averages and integrals over $\lambda$ \citep{mackenbachBounceaveragedDriftsEquivalent2023b}.

\section{Helmholtz free energy ($\Delta = 0$) bounds}
\begin{figure}
     \centering
     \begin{subfigure}[b]{0.5\textwidth}
         \centering
         \includegraphics[width = \textwidth]{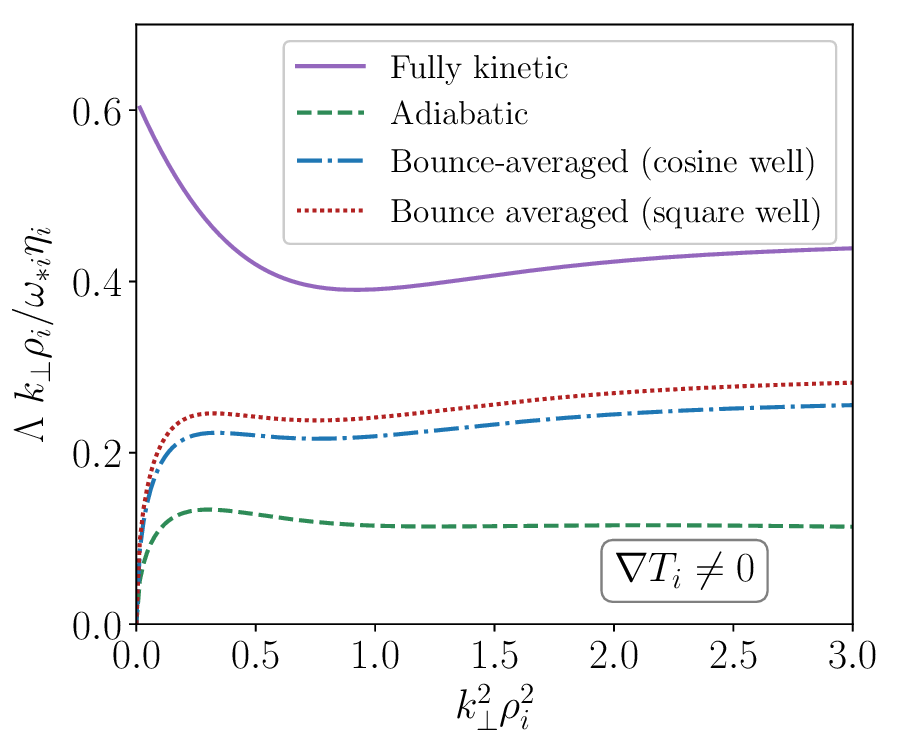}
         \caption{\protect\label{subfig:ITG_Helm}}
     \end{subfigure}%
     \begin{subfigure}[b]{0.5\textwidth}
         \centering
         \includegraphics[width = \textwidth]{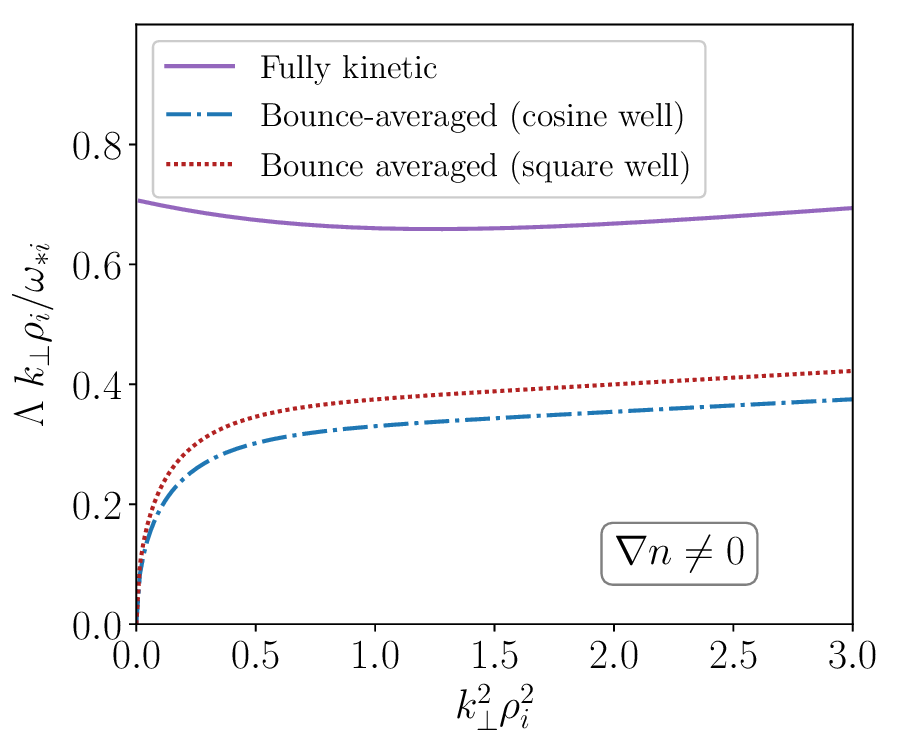}
         \caption{\protect\label{subfig:TEM_Helm}}
     \end{subfigure}
      \begin{subfigure}[b]{0.5\textwidth}
         \centering
         \includegraphics[width = \textwidth]{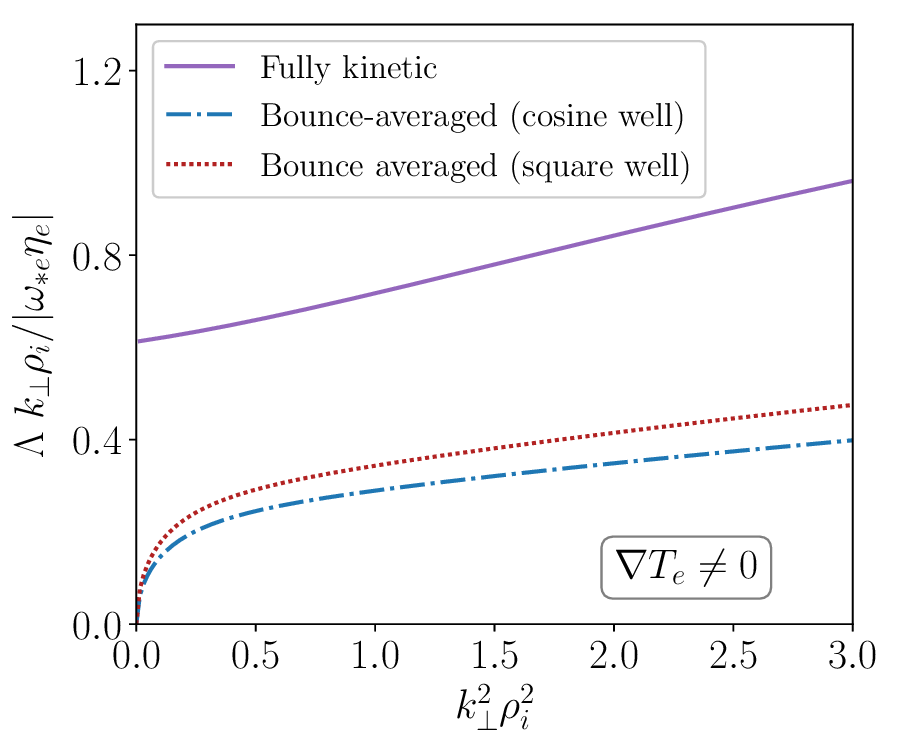}
         \caption{\protect\label{subfig:ETG_Helm}}
     \end{subfigure}
    \caption{Optimal mode growth rates $\Lambda$ of the Helmholtz free energy with $\tau= 1$ as a function of $k_\perp^2\rho_i^2$ for the various electron models. The top left figure shows the pure ITG case, and the top right shows the pure TEM case and the lower plot shows the pure ETG-TEM case.}
    \label{fig:Helmholt_bounds}
\end{figure}

\begin{figure}
    \centering
    \includegraphics[width=0.5\linewidth]{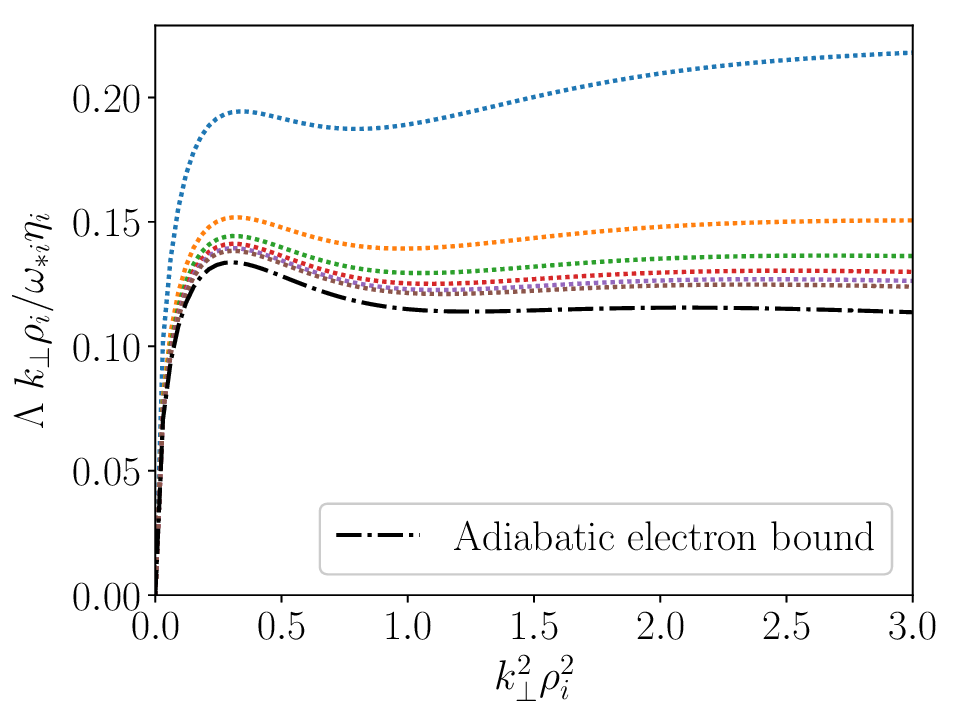}
    \caption{Spectrum of  numerical solutions to the kinetic eigenvalue problem \eqref{eqn:kinetic_eigenvalue_problem}} for a sinusoidal magnetic field strength with $\tau = 1$ for the pure-ITG case alongside the adiabatic electron optimal growth rate. Shown are the six largest eigenvalues found numerically.
    \label{fig:ITG_spectrum}
\end{figure}

\begin{figure}
    \centering
    \includegraphics[width=\linewidth]{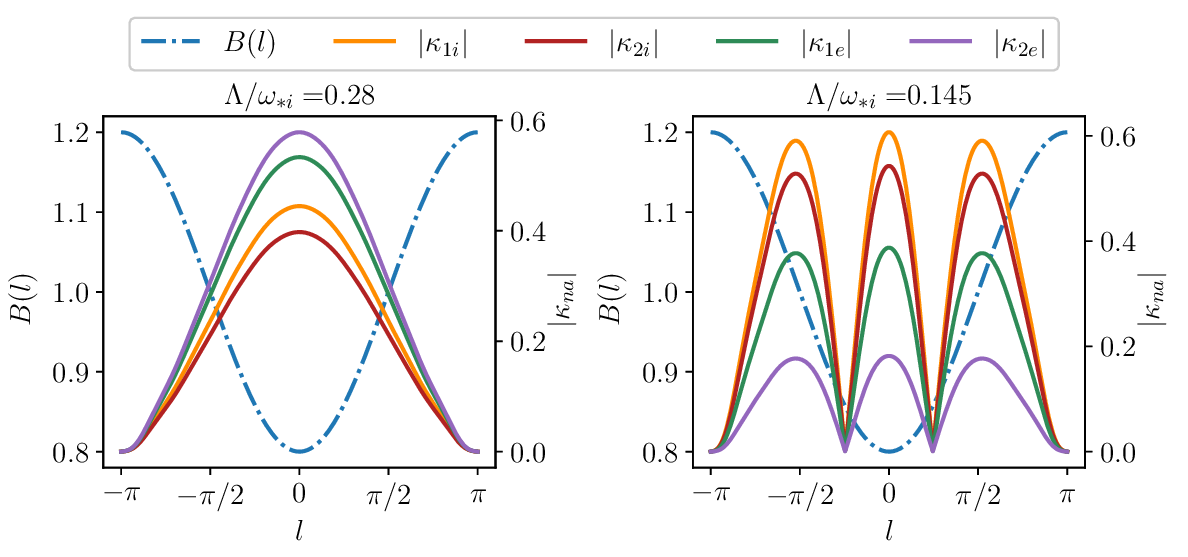}
    \caption{Eigenfunctions of the kinetic eigenvalue problem \eqref{eqn:kinetic_eigenvalue_problem}, obtained numerically,  for $\Delta = 0$ (Helmholtz limit of the generalised free energy) with $\tau = 1$, $b_i = 1.5$, in the pure TEM case. Here, the absolute values of the $\kappa_{na}$ moments are shown as a function of $l$, alongside the magnetic field strength $B(l)$, for the eigenfunction with the largest eigenvalue (left) and the second largest (right).}
    \label{fig:eigenfunctions_Helmholtz}
    \end{figure}
For the case of $\Delta = 0$, the generalised free energy reduces to the Helmholtz free energy, and the $\kappa_{3a}$ moments do not contribute to the system of equations. This reduces the dimensionality of the system to $4\times 4$. In this limit, the optimal modes do not depend on magnetic curvature, but retain dependency on $B(l)$ and the variation of $k_\perp$ with $l$. We will refer to the optimal mode growth rate of this system with $\Delta = 0$ as the \textit{Helmholtz bound}. To get a sense of how the optimal modes behave in response to the presence of different plasma gradients, we consider three cases: the case of a pure ion-temperature gradient in the plasma ($\nabla n = 0$, $\nabla T_e = 0$) that we refer to as the `pure ITG case', a pure density gradient ($\nabla T_i = 0$, $\nabla T_e = 0$) case that we refer to as the `pure TEM case', and a case with only an electron temperature gradient ($\nabla n = 0$, $\nabla T_i = 0$ which we refer to as the `pure ETG-TEM case'. While we have only considered single gradients here for simplicity, we note that the optimal modes can equally be applied to mixed gradient cases.

In Figure~\ref{fig:Helmholt_bounds}, we show the Helmholtz bound of the system as a function of $b_i = k_\perp^2\rho_i^2$. Here, for the numerical solution, we consider a magnetic field of the form $B(l) = B_0 - B_1 \cos(l)$ with $B_0 = 1$ and $B_1= 0.1$, and use $k_\perp = \mathrm{const.}$ for simplicity. In a sinusoidal field such as this, the magnetic trapping wells are identical and are separated by identical maxima (like in an omnigenous field), thus we need only solve the system in a single well with  $-\pi \leq l \leq \pi$. Note that numerically solving the integro-algebraic system results in a spectrum of eigenvalues (an example of which can be seen in  Figure \ref{fig:ITG_spectrum}), which are all real due to the Hermiticy of the system, and come in pairs of positive and negative values of equal magnitude due to the time reversibility of the system \citep{plunkEnergeticBoundsGyrokinetic2022}. For clarity, we only show the fastest-growing optimal mode growth rate (the largest positive eigenvalue).

We also show the analytical Helmholtz bound of the system in a square magnetic trapping well of the same depth, $\Lambda_{\mathrm{SW}}$, given by  \eqref{eqn:square_well_solution} with $\Delta = 0$. We compare these Helmholtz bounds with the bounce-averaged trapped electron response to the Helmholtz bounds with adiabatic electrons, given by equation (6.20) in \cite{helanderEnergeticBoundsGyrokinetic2022}, and the Helmholtz bound with fully kinetic electrons, taken from Appendix C of  \citet{plunkEnergeticBoundsGyrokinetic2022}.
For all gradients considered, we find that the numerical solution in the sinusoidal magnetic field closely follows the analytical solution for the square magnetic trapping well.

For the pure ITG case (top left of Figure~\ref{fig:Helmholt_bounds}), we find that the Helmholtz bound of the system with the bounce-averaged trapped electron response lies between the bounds for the adiabatic and fully-kinetic electron systems. Moreover, in Figure \ref{fig:ITG_spectrum} we see that this is true of the entire spectrum of solutions, which appear to be bounded from below in magnitude by the adiabatic electron bound.
We note that, for all gradients, the optimal mode growth rate tends to zero as $k_\perp \to 0$ in the bounce-averaged system. This signifies the removal of the `MHD-like' instability mechanism present in the fully kinetic electron Helmholtz bound, which causes the bound to approach a finite value in this limit. Mathematically, the cause of this difference is illustrated by the square well solution $\Lambda_{\mathrm{SW}}$ where as $k_\perp \to 0$, the numerator tends to zero and the denominator  is proportional to $(1 - \varepsilon)$, which is finite unless the trapped particle fraction is unity. Thus, we have $\Lambda_{\mathrm{SW}} \to $ 0 as $k_\perp \to 0$. 
This is in contrast to the fully kinetic electron bound of Part II, whose denominator goes to zero in the limit $k_\perp \to 0$ in such a way that the bound remains finite. Physically, the Helmholtz bound with fully-kinetic electrons over-estimates the degree of instability because it contains a contribution from non-adiabatic passing electrons which, in toroidal geometry, does not contribute on timescales longer than the electron transit time. 

The bounce-averaged electron system also yields a Helmholtz bound for the pure TEM case and the pure ETG-TEM case; see Figures \ref{subfig:TEM_Helm}--\ref{subfig:ETG_Helm}. In this bounce-averaged electron system, this provides a bound on the growth of $\nabla n$ and $\nabla T_e$-driven trapped electron modes (this does not extend to `true' ETGs, which exist at larger values of $k_\perp$ thus violating our assumption of $k_\perp \rho_e \ll 1$ and require a non-adiabatic passing electron response).


In Figure~\ref{fig:eigenfunctions_Helmholtz}, an example of two of the fastest-growing eigenmodes is shown for the pure TEM case. The eigenfunctions are similar (in $l$-variation) to the expectation from the linear theory, with the fastest growing mode having a maximum where the largest fraction of trapped electrons reside at the minimum of the magnetic field strength. The fastest-growing eigenmode has a significant contribution from the $\kappa_{ne}$ moments, indicating their importance for extracting free energy at these parameters.

\section{ Generalised free energy bounds}

\begin{figure}
    \centering
    \includegraphics[width=0.5\linewidth]{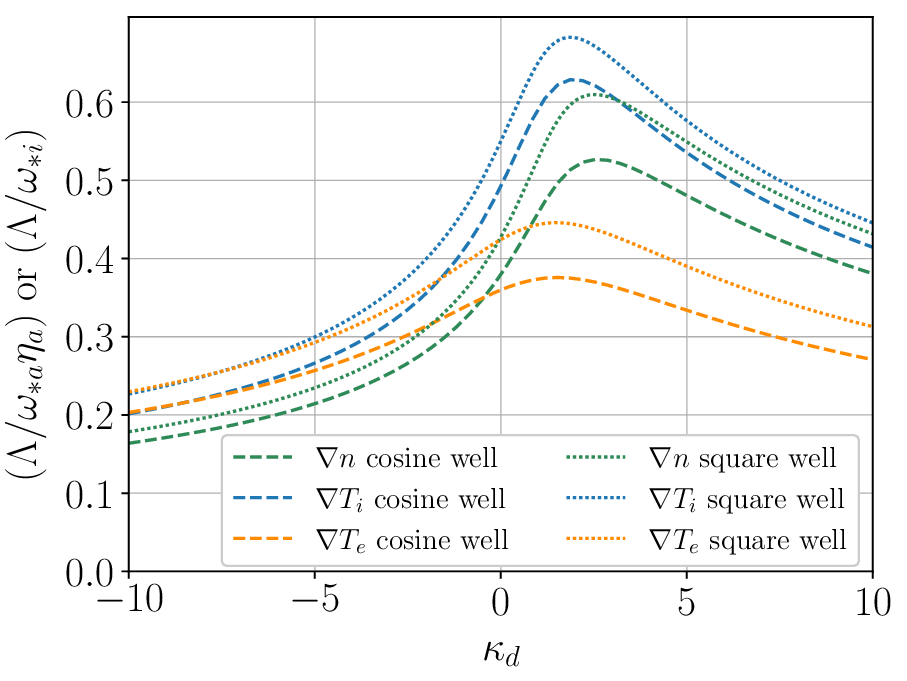}
    \caption{The optimal mode growth rate of generalised free energy, at the minimising value of $\Delta$, versus the drive parameter $\kappa_d$. The analytical square well with constant curvature solution is shown alongside the numerical solutions of the eigenvalue problem in a cosine magnetic well with constant curvature. These are computed for a density-gradient-driven case (pure TEM case), an ion-temperature-gradient-driven case (pure ITG case) and an electron-temperature-gradient-driven case (pure ETG-TEM case).}
    \label{fig:generalised_upper_bound_const_curvature}
\end{figure}

\begin{figure}
    \centering
    \includegraphics[width=0.5\linewidth]{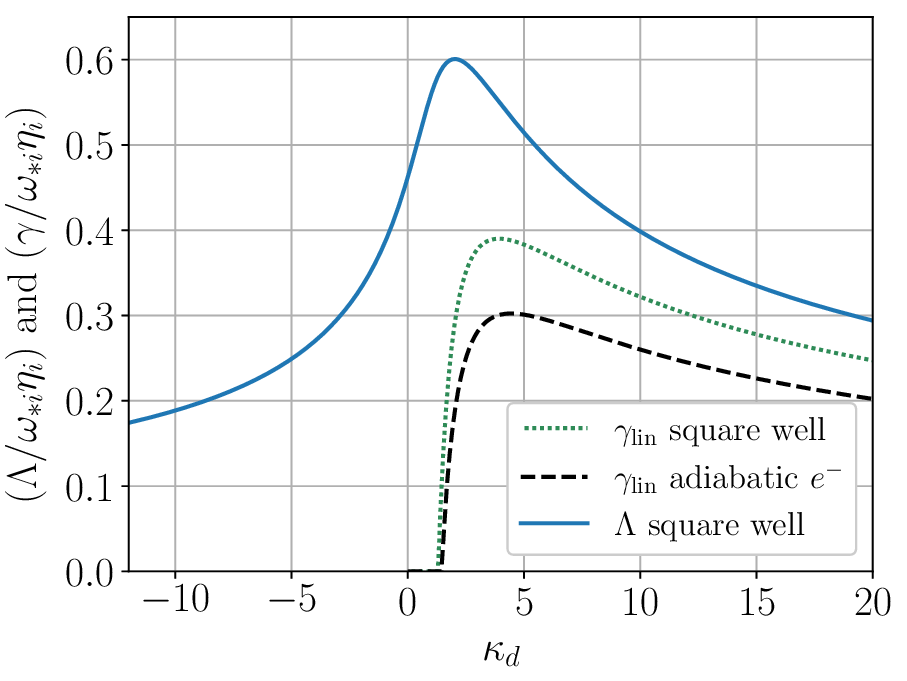}
    \caption{Comparison between the generalised upper bound  $\Lambda_{\mathrm{SW}}$, and the linear growth rate, for a pure ITG case in a square magnetic well with $B_{\mathrm{min}}/B_\mathrm{max} = 0.1$, $\tau = 1$, and constant curvature. Here, the drift-kinetic limit has been considered for simplicity ($J_{0i} \approx 1$).}
    \label{fig:comparison_with_linear}
\end{figure}

We now turn our attention to the general problem in which $\Delta$ is a free parameter. For this case where $\Delta \neq 0$, the species moments $\kappa_{3a}$, are retained. Thus, the optimal modes of this system depend on the magnetic curvature drift, $\omega_{da}$, for each species.  We seek values of $\Delta$ which minimise the maximum growth rate $\Lambda$ of the generalised free energy, giving the lowest possible upper bound.  The largest resulting optimal growth rate, at this minimising value of $\Delta$,  gives an upper bound on instabilities, in the considered limit, that depends on the magnetic curvature. 

Because of the complexity of the general system, we perform the optimisation over $\Delta$ numerically and exclude values for which positive definiteness is violated. However, we have found that the optimisation space of $\Lambda( \Delta)$ is very smooth and that a global minimum is easily found. 

As a first test for the numerical solution, to facilitate comparison with the analytical solution for square magnetic trapping well with constant curvature, we solve the system numerically for the case of a magnetic field of the form  $B(l) = B_0 - B_1 \cos(l)$ with $B_0 = 1$ and $B_1= 0.1$ with constant curvature, $\hat{\omega}_{da}(l) = \mathrm{const}$. Following the analysis of the generalised free energy with adiabatic electrons of Part III, we consider the instability drive parameter $\kappa_d = \omega_{*i}/\hat {\omega}_{di}$ or ($\kappa_d= \omega_{*i}\eta_i/\hat{\omega}_{di}$ for the pure ITG case). We thus have $\hat{\omega}_{da}/\omega_{*i} = \mathrm{sign}(e_a)\kappa_d^{-1}T_a/{T_i}$ for each species. Under this normalisation, if $\kappa_d > 0$, then $\hat\omega_{da}\omega_{*a} > 0$ for both species, corresponding to `bad' curvature where linear instability is expected. The case of $\kappa_d < 0$ corresponds to `good' curvature, where the system is expected to be linearly stable. 

In Figure~\ref{fig:generalised_upper_bound_const_curvature}, 
we show the optimal mode growth for the pure TEM case, the pure ETG-TEM case, and for the pure ITG case, as the curvature-drive parameter $\kappa_d$ is varied. We see that the square and the cosine trapping wells follow similar trends with $\kappa_d$, attaining a maximum for a finite positive value of $\kappa_d$, where the optimal bound is obtained for $\Delta = 0$ and therefore coincides with the Helmholtz bound. This feature of the generalised free energy bound overlapping with the Helmholtz bound at a particular value of the curvature drive was also found for the adiabatic electron analysis in Part III. The optimal bound decreases monotonically away from this maximum towards the `strongly driven' limit of large positive $\kappa_d$ and in the direction of small $\kappa_d$  in the `resonant limit'.  The bound also decreases for $\kappa_d \leq 0$ where the curvature is favourable (`good').

In Figure~\ref{fig:comparison_with_linear}, we compare the optimal mode growth in the square well, $\Lambda_{\mathrm{SW}}$, with constant curvature to the solution to the linear dispersion relation with bounce-averaged electrons (details given in Appendix~\ref{Appendix:linear_theory}) in the same geometry, in the pure ITG case. Once again we vary the drive parameter $\kappa_d$ to compare how each solution depends on the curvature inside the well. We confirm that the linear result lies below the upper bound, and we find that, much like the behaviour of the upper bound, the presence of bounce-averaged electrons in the linear dispersion relation `lifts up' the linear growth rate of the ITG. In regions where the normal modes are stable, the optimal modes have a finite but reduced growth rate, as was found in Part III for the adiabatic electron system.

\subsection{Impact of the maximum-$J$ property}

\begin{figure}
    \centering
    \includegraphics[width=0.75\linewidth]{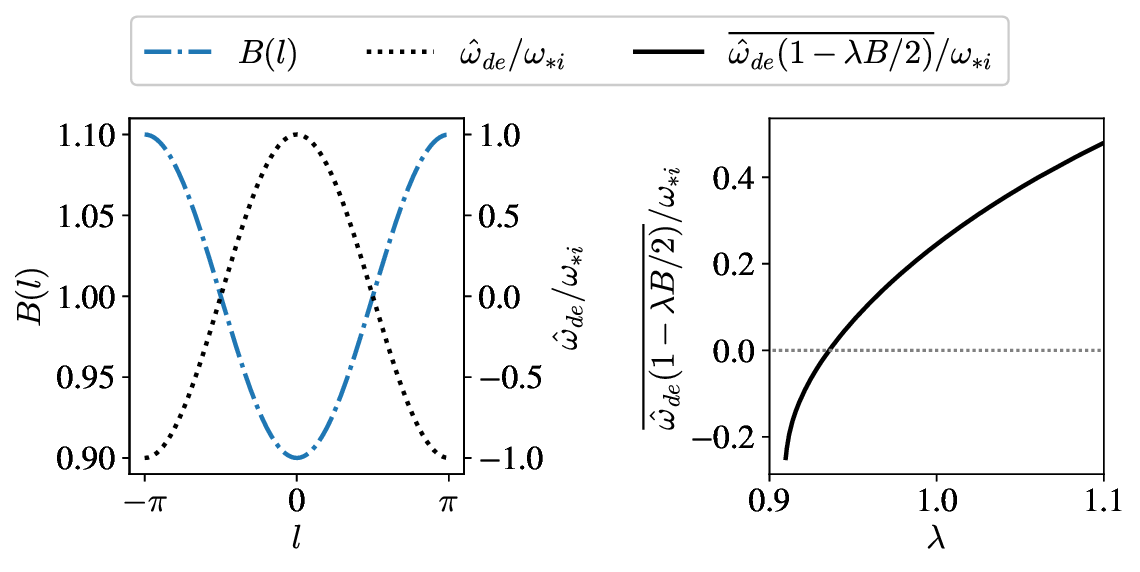}
    \caption{An example of an approximately maximum-$J$ toy geometry, where most trapped electrons, particularly those that are deeply trapped, experience bounce-averaged good curvature. In the left figure, the curvature is shown alongside the magnetic field strength, where negative values indicate bad curvature. On the right, the $\lambda$-dependence of the bounce-averaged electron drift is shown.}
    \label{fig:maximum-J-geometry}
\end{figure}

\begin{figure}
     \centering
     \begin{subfigure}[b]{0.5\textwidth}
         \centering
         \includegraphics[width = \textwidth]{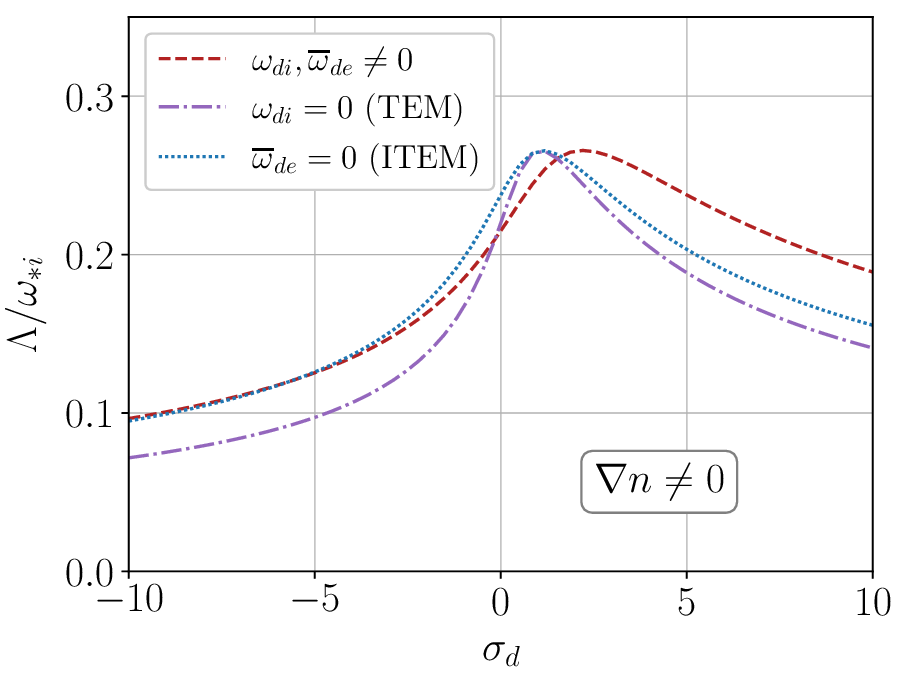}
     \end{subfigure}%
     \begin{subfigure}[b]{0.5\textwidth}
         \centering
         \includegraphics[width = \textwidth]{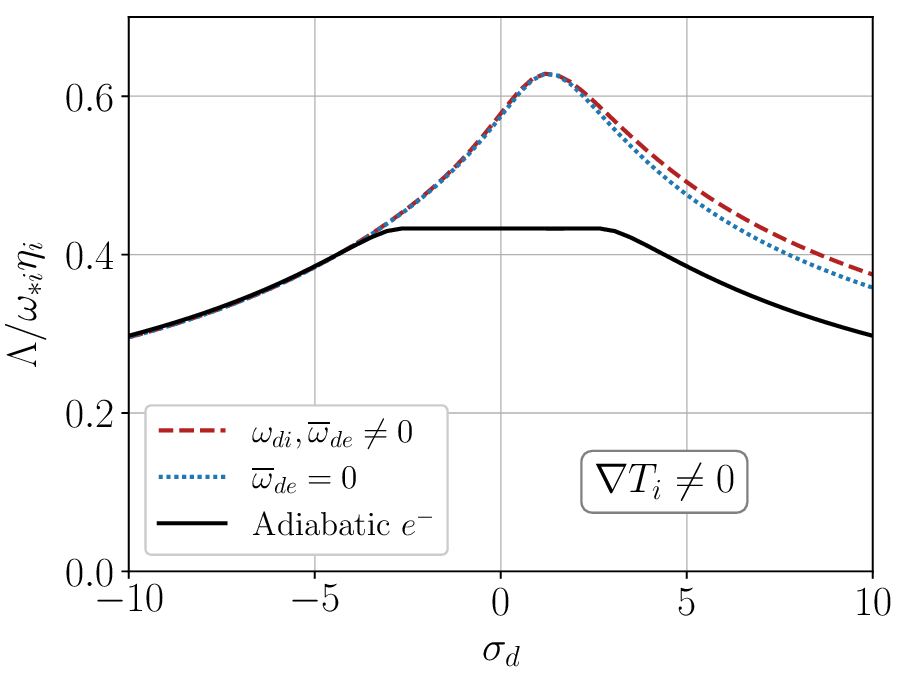}
         
     \end{subfigure}
          \begin{subfigure}[b]{0.5\textwidth}
         \centering
         \includegraphics[width = \textwidth]{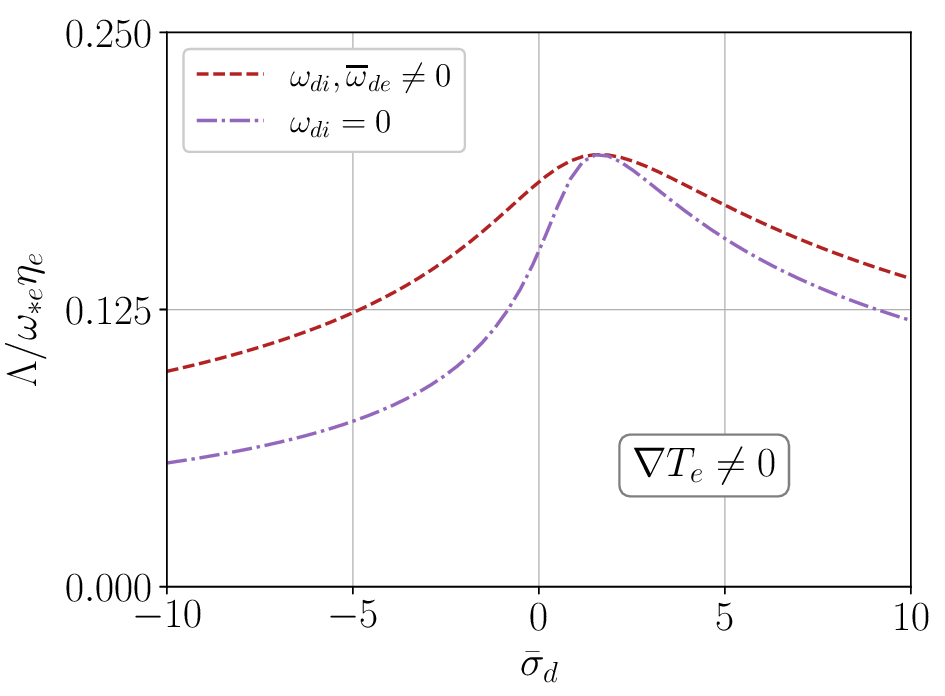}
         
     \end{subfigure}
         \caption{The optimal mode growth rate of generalised free energy, for at the minimising $\Delta$, versus the drive parameter $\sigma_d$. Negative and positive values of this parameter correspond to the `maximum-$J$' and `minimum-$J$' cases of the model geometry shown in Figure \ref{fig:maximum-J-geometry}. On the left, the pure TEM case is shown at $k_\perp \rho_i = 1.5$, on the right, the pure ITG case in the drift kinetic limit ($J_{0i} \approx 1$) is shown, and the lower plot shows the pue ETG-TEM case at $k_\perp\rho_i = 1.5$  The impact of species-dependent curvature drifts on each case is explored by zeroing out $\hat{\omega}_{di}$ or $\overline{\omega}_{de}$ independently.}
    \label{fig:max_J_vs_minimum_J}
\end{figure}

 A class of magnetic configurations in which the gyrokinetic system with bounce-averaged electrons is particularly interesting is provided by so-called `maximum-$J$' geometries, where the radial profile of the parallel adiabatic invariant,
 \begin{equation*}
     J(v,\lambda, \psi, \alpha) = m_e v\int_{l_1}^{l_2}\sqrt{1 - \lambda B} \ \mathrm{d} l,
 \end{equation*}
has a maximum at the magnetic axis, and decreases radially \citep{Rosenbluth-1968}. In such geometries, the electrons experience good curvature everywhere because
\begin{equation}
    \omega_{*e}\overline{\omega}_{de} = -\frac{T_e k_\alpha^2}{e^2\tau_B}\frac{\, \mathrm{d}\ln n}{\, \mathrm{d}\psi}\frac{\partial J}{\partial \psi}
\end{equation}
is negative for all trapped electrons if $\partial J / \partial \psi < 0$  and ${\, \mathrm{d}\ln n}/{\, \mathrm{d}\psi} < 0$, as is usually the case in fusion plasmas \citep{prollResilienceQuasiIsodynamicStellarators2012,helanderCollisionlessMicroinstabilitiesStellarators2013}. Quasi-isodynamic stellarators are a particular type of optimised stellarator \citep{Helander_2009,Nührenberg_2010} which can be designed to have this property \citep{rodriguezMaximumPropertyQuasiisodynamic2023,sanchezQuasiisodynamicConfigurationGood2023}.
In such devices, the local curvature (as seen by the ions) may be unfavourable, but the bounce-averaged curvature for trapped electrons is always favourable.  Many results from normal-mode theory, as well as gyrokinetic simulations, have found that the trapped electron population in these geometries have a stabilising influence if a plasma density gradient is present \citep{prollResilienceQuasiIsodynamicStellarators2012, plunkCollisionlessMicroinstabilitiesStellarators2014, prollTurbulenceMitigationMaximumJ2022,alcusonSuppressionElectrostaticMicroinstabilities2020}. The physical reason for this enhanced stability can be traced back to the fact that relatively little of the thermal energy is ``available'' for conversion to instabilities \citep{helanderAvailableEnergyGround2017,mackenbachAvailableEnergyTrapped2023}.

To investigate if this property is present in the optimal mode system, we require a test geometry where the ions may experience bad curvature locally, while the bounce-averaged curvature of the electrons can be independently varied. A simple model magnetic geometry can be seen in Figure~\ref{fig:maximum-J-geometry}. In this geometry, we choose a curvature of the form $\hat{\omega}_{da}(l)/\omega_{*i} = \mathrm{sign}(e_a)\sigma_d \cos(l) T_a/{T_i}$ and $B(l) = B_0 + B_1\cos(l)$, where once again, we take $B_0 = 1.0$ and $B_1 = 0.1$. In this toy geometry, $\sigma_d$ is a parameter which determines the relative magnitude of the gradient length scale and the maximum radius of curvature, similar to the familiar $R/L_n$ or $R/L_{T_a}$ normalisation, where $R$ is the major radius of the torus and $L_{n,T_i}$ is the scale length of the density or temperature gradients, respectively.
With this model geometry, by changing the sign of $\sigma_d$ we can choose between a mostly maximum-$J$ geometry for $\sigma_d < 0$, where most trapped electrons, particularly those which are deeply trapped, experience bounce-averaged good curvature, and a mostly minimum-$J$ geometry for $\sigma_d > 0$. Importantly, we can make this change between maximum-$J$ and minimum-$J$ geometry without changing the field-line-average curvature (zero in this case), whilst always having a region of unfavourable curvature along the field line.

In Figure~\ref{fig:max_J_vs_minimum_J}, we show the optimal growth rate $\Lambda$ as $\sigma_d$ is varied, the pure ITG case, pure TEM case and pure ETG-TEM case.  The upper bound in the pure TEM case is lower in the maximum-$J$ geometries (negative  $\sigma_d$) than in the equivalent minimum-$J$ geometries (positive $\sigma_d$). To explore the physical mechanism at play, we also test the dependence of the optimal mode growth on the curvature drift for each species separately. For $\sigma_d<0$ (maximum-$J$), removing the electron drift frequency (the instability mechanism that remains is analogous to the ITEM of \citet{plunkCollisionlessMicroinstabilitiesStellarators2017}) increases growth rates if $|\sigma_d|$ is small, and has little effect for $\sigma_d$ large and negative. Thus, $\overline{\omega}_{de}$ has a stabilising effect in the maximum-$J$ geometries, in agreement with the expectation linear theory. Moreover, in maximum-$J$ geometries at large  positive values of $\sigma_d$, the instability mechanism is largely reliant on the ion curvature drift, as is expected from the linear ITEM or ubiquitous mode \citep{coppiTheoryUbiquitousMode1977}.

 For $\sigma_d > 0$ (minimum-$J$), the removal of the curvature drift for either species in the pure TEM case serves to move the maximum value of the optimal growth rate towards lower values of $\sigma_d$ and greatly reduces the growth rate for large $\sigma_d$. This suggests that the drifts of both species are important for the optimal extraction of free energy from the gradients. 

 In the pure ETG-TEM case, we see a reduction of the optimal mode growth for $\sigma_d < 0$. This is in keeping with the expectation from the dispersion relation of \citet{plunkCollisionlessMicroinstabilitiesStellarators2017}, which shows that the maximum-$J$ property, associated with the relative direction of the bounce-averaged drift to the electron diamagnetic frequency, also exerts a stabilising influence in the strongly driven limit when $\eta_e$ is finite. The optimal mode analysis here suggests that this benefit extends into the resonant limit (small $\sigma_d$).
 
 We observe that maximum-$J$ configurations also exhibit lower optimal growth in the ITG scenario ($\nabla n = 0$, $\nabla T_e = 0$). This is despite the absence of any electron free-energy source, something which has not been predicted previously from linear normal-mode theory.  It is found that beyond a negative value of $\sigma_d$ of sufficient magnitude, the ITG upper bound with bounce-averaged electrons coincides with that of the adiabatic electron upper bound of Part III computed in this geometry (because the bound with adiabatic electrons is local in this limit, here we evaluate the bound at each point in $l$ in the toy geometry and plot the largest value found). This is also reflected in the eigenfunction (not shown), which tends towards a delta function at the location of  maximum magnetic field strength, where the trapped electron contribution is zero, suggesting that the adiabatic ITG solution to \eqref{eqn:kinetic_eigenvalue_problem} becomes the solution with the largest $\Lambda$ for these values of $\sigma_d$. As shown in Figure~\ref{fig:max_J_vs_minimum_J}, this mechanism of stabilisation in maximum-$J$ geometry is present even if  $\overline{\omega}_{de} = 0$.  Thus, this stabilising mechanism in maximum-$J$ geometries is distinct from the known mechanism that follows from the difference in sign between $\overline{\omega}_{de}$ and $\omega_{*e}^T$ \citep{prollTurbulenceMitigationMaximumJ2022}.
\subsubsection{Interpreting with linear theory}
To interpret the observed reduction of the ITG-driven optimal mode growth in maximum-$J$ geometries without a source of electron free energy, we turn to expressions from linear normal-mode theory (i.e, solutions to the gyokinetic equation which evolve in time like $\exp({-i \omega t})$, where $\omega$ is the complex freqeuncy). The comparable linear dispersion relation is given by equation (3.5) in \citet{plunkCollisionlessMicroinstabilitiesStellarators2017}, for the case of a pure ITG, which is valid if  the mode frequency satisfies $\omega/(\omega_{*i}\eta_i) \sim (\omega_{di}/\omega_{*i}\eta_i)^{1/2} \ll 1$ (i.e. in the regime of large $\kappa_d$). Taking the drift-kinetic limit ($J_{0i} \approx 1)$, the normal mode frequency is given by,
\begin{multline}
\label{eqn:linear_ITG_growth_rate_plunk_2017}
    \omega = \pm \sqrt{-\omega_{*i}\eta_i \int_{-\infty}^{+\infty} \hat{\omega}_{di}(l) | \delta \phi|^2 \frac{\mathrm{d}l}{B}} \Bigg/  \\
    {\left(\tau \int_{-\infty}^{+\infty}| \delta \phi|^2 \frac{\mathrm{d}l}{B} - \frac{\tau}{2}\sum_j \int_{1/B_{\mathrm{min}}}^{1/B_{\mathrm{max}}} \tau_{B,j} |\overline{\delta \phi}_{j} |^2\, \mathrm{d} \lambda\right)^{\frac{1}{2}}},
\end{multline}
where instability growth is given by a positive imaginary part of $\omega$ which arises due to regions of bad curvature along the magnetic field ($\omega_{*i}\eta_i \hat\omega_{di} > 0$). We see that when the ITG is unstable, the trapped electron contribution in the denominator acts to further destabilise the mode
by making the denominator (which is always positive \citep{helanderCollisionlessMicroinstabilitiesStellarators2013}) smaller. This destabilising effect is due to the depletion of the Boltzmann response by the population of non-adiabatic trapped electrons.

The effect of the maximum-$J$ property on this dispersion relation is most easily illustrated when the case of the `square well' magnetic field with constant curvature inside the well is considered. Inside the well, the ITG frequency becomes $ \omega = \pm \sqrt{-\omega_{*i}\eta_i \hat{\omega}_{di}/ (\tau\left(1- \varepsilon\right))}$,
where $\varepsilon$ is the trapped particle fraction, and outside the well it is given by  $\omega = \pm \sqrt{-\omega_{*i}\eta_i \hat{\omega}_{di}/\tau}$.
If we consider a series of these square magnetic trapping wells with minima of $B = B_{\mathrm{min}}$ separated by regions of $B = B_{\mathrm{max}}$, with a magnetic curvature that alternates sign (from `good' curvature to `bad' curvature) between regions of $B = B_{\mathrm{max}}$  and $B = B_{\mathrm{min}}$, then there are two possibilities: either the bad curvature is located at the minimum of magnetic field strength, corresponding to a minimum-$J$ geometry, or the bad curvature is located outside the well, corresponding to a maximum-$J$ geometry.

In the minimum-$J$ case the ITG is unstable \textit{inside} the well, with a growth rate given by $\omega = \pm \sqrt{-\omega_{*i}\eta_i \hat{\omega}_{di}/ (\tau\left(1- \varepsilon\right))}$. In this scenario, the non-adiabatic population of trapped electrons can raise the ITG growth rate by depleting the Boltzmann response. On the other hand, in the maximum-$J$ case, the ITG is stable inside the well because the curvature is favourable there ($\hat{\omega}_{di} \omega_{*i}\eta_i> 0$). Instead the unstable ITG is \textit{outside} of the well at $B= B_{\mathrm{max}}$ and the growth rate is  $\omega = \pm \sqrt{-\omega_{*i}\eta_i \hat{\omega}_{di}/\tau}$, corresponding to the ITG growth rate if the electrons are treated adiabatically. 

This behaviour can also be inferred from the ITG growth rate in general geometry \eqref{eqn:linear_ITG_growth_rate_plunk_2017}, in which it can be seen that for instability to occur, the average of $\omega_{*i}\eta_i\hat\omega_{di}| \delta \phi|^2$ along the magnetic field line must be positive, such that $\delta \phi$ must have large amplitude in the regions of bad curvature.
In maximum-$J$ geometry the regions of bad curvature are at different locations in $l$ than the minima of magnetic field strength. In these geometries, for the instability criterion to be fulfilled, the non-adiabatic electron response, which is proportional the the average potential seen by the trapped electrons as they bounce in the magnetic well, will be necessarily small. Otherwise, if the electron contribution were large, the potential would have a large amplitude in a region of favourable curvature, thus impacting the instability criterion. As a result, the destabilising impact of trapped electrons on the ITG is diminished in such geometries.

This is not the case in minimum-$J$ geometries, where the regions of bad curvature are located at the minima of $B$ along the field line. In these geometries, the instability criterion and the magnitude of the trapped electron contribution are not in conflict, thus the ITG growth can be increased by the trapped electron population without impacting the instability mechanism.

In the optimal mode analysis, this effect is slightly obscured due to the optimisation over the $\Delta$ parameter, but the same basic behaviour can be observed in the expression for $E(\mathbf{k}, t)$ given by \eqref{eqn:electrostatic_energy}, which is of the same form as the denominator in \eqref{eqn:linear_ITG_growth_rate_plunk_2017}. Thus, the magnitude of $E(\mathbf{k},t)$ affects the magnitude of $\Lambda$ via the denominator of \eqref{eqn:Lamda_definition}, for  non-zero values of $\Delta$. This stabilising effect of maximum-$J$ can be seen in Figure \ref{fig:max_J_vs_minimum_J}, where for the ITG scenario, the upper bound with bounce averaged electrons approaches the adiabatic electron result for large negative values of $\sigma_d$, corresponding to an approximately maximum-$J$ geometry.

To summarise this section, the impact of the trapped electron response on ITG instability in the absence of electron free energy is always destabilising, but the amount of destabilisation is impacted by the location of the most deeply trapped particles relative to the regions of bad curvature.  To use a term coined in the literature for zonal-flows \citep{diamondZonalFlowsPlasma2005}, the location of the trapped particles changes the \textit{inertia} of the ITG mode, with ITGs possessing lower inertia in minimum-$J$ configurations than in those which are maximum-$J$. This may go some way towards explaining why simulations performed in newly optimised, highly-maximum-$J$ stellarators \citep{goodmanQuasiisodynamicStellaratorsLow2024a}  show a smaller increase in ITG-driven heat fluxes with the addition of kinetic electrons in comparison to configurations which do not satisfy the maximum-$J$ property to the same degree.

\section{Conclusion}
We have studied the optimal modes of the generalised free energy  for the case of a two-species system with fully gyrokinetic ions and bounce-averaged electrons. The central result of this effort is a system of integro-differential equations \eqref{eqn:1stionmoment}--\eqref{eqn:3rdionmoment} and \eqref{eqn:1stelectronmoment}--\eqref{eqn:3rdelectronmoment}, the solutions to which provide an upper bound on the possible instantaneous growth of any instabilities which satisfy $L/v_{Te} \ll \tau_D$. These bounds depend explicitly on the magnetic field strength of the chosen flux tube.
Despite being more difficult to solve than the optimal mode systems of Parts I-III (requiring a numerical treatment for all but the simplest of magnetic fields), the eigenvalue problem here, which involves a relatively small number of velocity space moments of the distribution function, is still much simpler than the equivalent normal mode problem, which involves computing the full distribution function as a function of velocity space. 

As is discussed in Parts I-III, the upper bounds given by the fastest growing optimal modes not only give an upper bound on linear instability, but also have nonlinear implications. This is because the nonlinear growth of the system is bounded by the growth of the fastest growing optimal mode. Here, however, the removal of the `finite-electron-Larmor-radius' effects associated with retaining $k_\perp\rho_e$ results in an unbounded growth of $\Lambda$ as $k_\perp \to \infty$. As a result, the nonlinear growth is also unbounded if all values of $k_\perp$ are considered. Thus, when seeking a bound on nonlinear growth, the spectrum in $k_\perp$ must be cut off at a chosen value, limiting its applicability to cases in which only the ion-scale dynamics are of interest, as is often done in gyrokinetic simulations.
Other nonlinear implications of optimal modes also discussed in Part II, such as their required presence in statistically steady turbulence \citep{landremanGeneralizedUniversalInstability2015, delsoleNecessityInstantaneousOptimals2004}, also hold true for the optimal modes studied here.


The upper bounds computed here are specific to toroidal geometries with non-periodic flux tubes (i.e tokamaks and stellarators) and as such, the Helmholtz bound we find behaves as expected in these geometries at small values of $k_\perp\rho_i$. This is in contrast to the fully-kinetic-electron Helmholtz bound of Part II which, in its generality, must also bound growth in closed-field-line geometries. The Helmholtz bounds of the bounce-averaged system are significantly lower than the Part II result for all gradients, as long as $k_\perp \rho_e \ll 1$.

Moreover, the generalised bound, which is optimised over the free parameter $\Delta$, depends explicitly on all the geometry quantities associated with the flux-tube domain. Here, we neglected parallel ion motion in our system of equations for the sake of simplicity, focusing on the typical toroidal ITG/TEM limits familiar from normal-mode theory. We found that the optimal modes exhibit much of the behaviour expected from normal modes, attaining a maximum growth rate (relative to the strength of the plasma gradients) when the curvature drift for both species is unfavourable and when the ratios of the diamagnetic frequencies and drift frequencies are close to unity. We have also demonstrated that the beneficial properties of maximum-$J$ configurations expected from normal-mode theory are also captured by the optimal modes. This was also found to be the case for another measure of turbulent transport that is valid nonlinearly, the \textit{available energy} of trapped electrons \citep{helanderAvailableEnergyGround2017, mackenbachAvailableEnergyTrapped2023}.

An unexpected consequence of this work was the observation of reduced ITG mode growth rates in maximum-$J$ configurations in comparison to a similar minimum-$J$ configuration when trapped electrons are included, even in the absence of a source of free energy in the electron equation. This effect, observed in both the optimal modes and the normal modes, was found to be related to the `inertia' of the ITG mode, which is larger in maximum-$J$ devices. 


The optimal modes presented in this work could form the basis of a `target function' for stellarator optimisation, given that they depend explicitly on the magnetic geometry and plasma parameters.  The optimal modes studied here exhibit the effects present in the adiabatic electron optimal modes of Part III. In particular, the two-species upper bound here shows a reduction of instability in both the `strongly-driven' and `resonant' limits of the drive parameter $\kappa_d$ (or $\sigma_d$); these limits are also visible in the available energy analysis of \citet{mackenbachAvailableEnergyTrapped2023}. This suggests that, for fixed plasma gradients, one could choose to reduce the amount of bad curvature experienced by both species thus increasing $\kappa_d$ (or $\sigma_d$). Alternatively, in so-called `critical-gradient optimisation', the amount of bad curvature could be increased (decreasing $\kappa_d$ into the resonant limit), as performed by \citet{roberg-clarkCriticalGradientTurbulence2023} for ITG-driven turbulence with adiabatic electrons. Our findings here suggest that both strategies are viable for reducing the growth of instabilities including the response of trapped electrons. Moreover, as we have seen, these optimal modes capture the interaction of several physical effects associated with the maximum-$J$ property and thus may directly inform the optimisation of the benefits of this property in a way that is not captured by present maximum-$J$ targets, which are generally not designed with turbulence in mind.

\section*{Funding}
This work has been carried out within the framework of the EUROfusion Consortium, funded by the European Union via the Euratom Research and Training Programme (Grant Agreement No 101052200 -- EUROfusion). Views and opinions expressed are however those of the author(s) only and do not necessarily reflect those of the European Union or the European Commission. Neither the European Union nor the European Commission can be held responsible for them. This work was partly supported by a grant from the Simons Foundation (560651, PH).

\section*{Declaration of interests}
The authors report no conflict of interest.

\appendix
\section{Determining the Hermitian linear operators}
\label{appendix:Hermitian_linear_ops}
The Hermitian linear operators in \eqref{eqn:kinetic_eigenvalue_problem}  can  be identified from the variational problem \eqref{eqn:variational_problem}. To make the variation of $\tilde{H}$ more straightforward we invert the identity \eqref{eqn:dlambda_dl_identity} to arrive at,
\begin{multline}
        \tilde{H} = \Bigg\langle T_e \int \frac{|g_{e}|^2}{F_{e0}} \, \mathrm{d}^3 v  + T_i \int \frac{|g_{i}|^2}{F_{i0}}\, \mathrm{d}^3 v  \\ - \frac{e^2 n}{T_i}\left(\alpha |\delta \phi|^2 - \frac{\Delta \tau}{2} \delta \phi^* \int_{1/B_\mathrm{max}}^{1/B(l)}\frac{B \, \mathrm{d}\lambda}{\sqrt{1 - \lambda B}} \overline{\delta \phi}\right) \Bigg\rangle,
\end{multline}
Where we have defined $\alpha = 1+ \tau + \Delta( 1 + \tau - \Gamma_{0i})$. Written in this form, all that remains is to insert $\delta \phi$ from quasi-neutrality,
\begin{equation}
    \delta \phi = \frac{T_i}{e n (1 + \tau)}\left(\int g_{i} J_{0i} \, \mathrm{d}^3 v -  \int g_{e} \, \mathrm{d}^3 v  \right),
\end{equation}
and perform the variation over $g_i$ and $g_e$, writing the result in terms of the inner product defined for each species given by \eqref{eqn:electron_inner_prod} and \eqref{eqn:ion_inner_prod} giving
\begin{equation}
    \frac{\delta \tilde{H}}{\delta g_{i}} = \left\langle \int \, \mathrm{d}^3 v \frac{T_i}{F_{i0}} g_{i}^* \left\{ \sum_b \tilde{\mathcal{H}}_{ib} g_{b}  \right \} \right\rangle + \mathrm{c.c},
\end{equation}
and
\begin{equation}
    \frac{\delta \tilde{H}}{\delta g_{e}} = \frac{2\pi}{V}\sum_{j}\int_{1/B_{\mathrm{max}}}^{1/B_{\mathrm{min}}} \, \mathrm{d} \lambda \int_{0}^{\infty} \, \mathrm{d} v  v^2 \tau_{B,j} \frac{T_e}{F_{e0}} g_{e}^* \left\{\sum_b \tilde{\mathcal{H}}_{eb} g_{b} \right\} + \mathrm{c.c},
\end{equation}
from which the operators can be identified. The construction of the operators associated with $D$ proceeds similarly. However, due to the derivatives with respect to $l$, care must be taken when finding the operators associated with $K$.  This involves integration by parts, and because the derivatives with $l$ are taken at constant ${E}_a$ and $\mu_a$,  we require the identity given in an appendix of \citet{plunkEnergeticBoundsGyrokinetic2023a},
\begin{equation}
    \int v_\parallel \frac{\partial g_{i}}{\partial l} J_{0i}\,\, \mathrm{d}^3 v = B\frac{\partial}{\partial l } \left(\frac{1}{B}\int v_\parallel g_{i}\, \mathrm{d}^3 v\right) - \int v_\parallel g_{i} \frac{\partial J_{0i}}{\partial l}\, \mathrm{d}^3 v.
\end{equation}
After some manipulation, the Hermitian linear operators can be found to be:
\begin{equation}
    \sum_b\tilde{\mathcal{H}}_{i b}g_b = g_i -\frac{F_{i0}J_{0i}}{n(1 + \tau)^2}\left(\alpha(\kappa_{1i} - \kappa_{1e}) -  \frac{\Delta \tau}{2}\int_{1/B_\mathrm{max}}^{1/B(l)}\frac{B \, \mathrm{d}\lambda}{\sqrt{1 - \lambda B}}(\overline{\kappa}_{1i} - \overline{\kappa}_{1e})\right),
\end{equation}
\begin{equation}
    \sum_b\tilde{\mathcal{H}}_{e b}g_b = g_e +\frac{\tau F_{e0}}{n(1 + \tau)^2}\left(\overline{\alpha(\kappa_{1i} - \kappa_{1e})} -  \frac{\Delta \tau}{2}\overline{\int_{1/B_\mathrm{max}}^{1/B(l)}\frac{B \, \mathrm{d}\lambda}{\sqrt{1 - \lambda B}}(\overline{\kappa}_{1i} - \overline{\kappa}_{1e})}\right),
\end{equation}
\begin{multline}
\sum_b \mathcal{D}_{ib}g_b = \frac{iF_{i0}J_{0i}}{2n(1+ \tau)}\Bigg(\omega_{*i}\left(1 + \eta_i\left(\frac{v^2}{v_{Ti}^2} - \frac{3}{2}\right )\right)(\kappa_{1i} - \kappa_{1e}) \\
- \omega_{*i}(1 - 3\eta_i/2)\kappa_{1i}  - \omega_{*i}\eta_i \kappa_{2i} + 
\omega_{*e}(1 - 3\eta_e/2)\kappa_{1e} + \omega_{*e}\eta_e\kappa_{2e} \Bigg),
\end{multline}
\begin{multline}
\sum_b \mathcal{D}_{eb}g_b = -\frac{i\tau F_{e0}}{2n(1+ \tau)}\Bigg(\omega_{*e}\left(1 + \eta_e\left(\frac{v^2}{v_{Te}^2} - \frac{3}{2}\right )\right)(\overline{\kappa}_{1i} - \overline{\kappa}_{1e}) \\
+ \omega_{*e}(1 - 3\eta_e/2)\overline{\kappa}_{1e}  + \omega_{*e}\eta_e \overline{\kappa}_{2e} - 
\omega_{*i}(1 - 3\eta_i/2)\overline{\kappa}_{1i} - \omega_{*i}\eta_i\overline{\kappa}_{2i} \Bigg),
\end{multline}
\begin{multline}
    \sum_b \mathcal{K}_{ib}g_b = \frac{F_{i0}}{2 n (1 + \tau)}\bigg( -J_{0i}B \frac{\partial}{\partial l }\left(\frac{v_{Ti}}{B}\kappa_{4 i} \right) + J_{0i}\kappa_{5i} v_{Ti } + v_\parallel \frac{\partial}{\partial l}\left (J_{0i}(\kappa_{1i}- \kappa_{1e})\right ) \\
    -iJ_{0i}\bigg[ \hat{\omega}_{di}(l)\kappa_{3i} -\tilde{\omega}_{de}\kappa_{3e}- \hat{\omega}_{di}(l)\left( \frac{v_\perp^2}{2 v_{Ti}^2} + \frac{v_\parallel^2}{v_{Ti}^2} \right)(\kappa_{1i} - \kappa_{1e})\bigg]\bigg),
\end{multline}
\begin{multline}
    \sum_b \mathcal{K}_{eb}g_b = \frac{\tau F_{e0}}{2 n (1 + \tau)}\bigg(\overline{B\frac{\partial}{\partial l}\left(\frac{v_{Ti}}{B}\kappa_{4i} \right)} - v_{Ti}\overline{\kappa_{5i}} + i\left[\overline{\hat{\omega}_{di}(l)\kappa_{3i}} + \tilde{\omega}_{de}\overline{\kappa}_{3e}\right]\bigg),
\end{multline}
where we have used the moment representation given by \eqref{eqn:moment_defs_1}--\eqref{eqn:moment_defs_2}. Inserting these operators into \eqref{eqn:kinetic_eigenvalue_problem}, for $a = i$ and $a = e$, gives the equation for the optimal distribution function for ions and electrons, respectively.

\section{Moment form of eigenvalue problem}
\label{appendix:Moment_forms}
Here, we give the (quite lengthy) closed set of equations which arise upon taking moments of \eqref{eqn:kinetic_eigenvalue_problem}. The moment equations which come from the optimal ion distribution function equation are
\begin{multline}
\label{eqn:1stionmoment}
    2\Lambda\left((1 + \tau)\kappa_{1i} - \frac{G_{0}}{(1 + \tau)}\left\{ \alpha (\kappa_{1i} - \kappa_{1e}) - \frac{\Delta\tau}{2}\int_{1/B_\mathrm{max}}^{1/B(l)} \frac{B(l) \, \mathrm{d}\lambda}{\sqrt{1 - \lambda B}} (\overline{\kappa_{1i}}- \overline{\kappa_{1e}}) \right\} \right) \\
    =i\bigg(\omega_{*i}\eta_i G_{1}(\kappa_{1i} -\kappa_{1e}) - \omega_{*i}\eta_i G_{0}\kappa_{2i} -\omega_{*i}(1 - 3\eta_i/2 )G_{0}\kappa_{1e} + \omega_{*e}\eta_e G_{0}\kappa_{2e} \\
    + \omega_{*e}(1 - 3\eta_e/2)G_{0}\kappa_{1e}  - \Delta\bigg\{\hat{\omega}_{di}(l)G_{3}(\kappa_{1i} - \kappa_{1e})+ \tilde{\omega}_{de}G_{0}\kappa_{3e} - \hat{\omega}_{di}(l)G_{0}\kappa_{3i} \bigg\}\bigg),
\end{multline}
\begin{multline}
\label{eqn:2ndionmoment}
    2\Lambda\left((1 + \tau)\kappa_{2i} - \frac{G_{1}}{(1 + \tau)}\left\{ \alpha (\kappa_{1i} - \kappa_{1e}) - \frac{\Delta\tau}{2}\int_{1/B_\mathrm{max}}^{1/B(l)} \frac{B(l) \, \mathrm{d}\lambda}{\sqrt{1 - \lambda B}} (\overline{\kappa_{1i}}- \overline{\kappa_{1e}}) \right\} \right) \\
    =i\bigg(\omega_{*i}\eta_i G_{2}(\kappa_{1i} -\kappa_{1e}) - \omega_{*i}\eta_i G_{1}\kappa_{2i} -\omega_{*i}(1 - 3\eta_i/2 )G_{1}\kappa_{1e} + \omega_{*e}\eta_e G_{1}\kappa_{2e} \\
    + \omega_{*e}(1 - 3\eta_e/2)G_{1}\kappa_{1e}   - \Delta\bigg\{\hat{\omega}_{di}(l)G_{4}(\kappa_{1i} - \kappa_{1e})+ \tilde{\omega}_{de}G_{1}\kappa_{3e} - \hat{\omega}_{di}(l)G_{1}\kappa_{3i} \bigg\}\bigg),
\end{multline}
\begin{multline}
\label{eqn:3rdionmoment}
    2\Lambda\left((1 + \tau)\kappa_{3i} - \frac{G_{3}}{(1 + \tau)}\left\{ \alpha (\kappa_{1i} - \kappa_{1e}) - \frac{\Delta\tau}{2}\int_{1/B_\mathrm{max}}^{1/B(l)} \frac{B(l) \, \mathrm{d}\lambda}{\sqrt{1 - \lambda B}} (\overline{\kappa_{1i}}- \overline{\kappa_{1e}}) \right\} \right) \\
    =i\bigg(\omega_{*i}\eta_i G_{4}(\kappa_{1i} -\kappa_{1e}) - \omega_{*i}\eta_i G_{3}\kappa_{2i} -\omega_{*i}(1 - 3\eta_i/2 )G_{3}\kappa_{1e} + \omega_{*e}\eta_e G_{3}\kappa_{2e} \\
    + \omega_{*e}(1 - 3\eta_e/2)G_{3}\kappa_{1e}    - \Delta\bigg\{\hat{\omega}_{di}(l)G_{5}(\kappa_{1i} - \kappa_{1e})+ \tilde{\omega}_{de}G_{3}\kappa_{3e} - \hat{\omega}_{di}(l)G_{3}\kappa_{3i} \bigg\}\bigg).
\end{multline}
Here, the functions $G_{ni} = G_{ni}(b)$ come from evaluating the integrals over velocity space which contain $J_{0i}^2$ in the integrand. We use the same notation as was used in Part III, where, in an Appendix, these functions were computed and are given as
\begin{align}
  G_{0} &= \Gamma_{0},\\ 
       G_{1} &= \left(\frac{3}{2}-b\right) \Gamma_0+b \Gamma_1, \\
         G_{2} &= \frac{1}{4} \left(\left(6 b^2-20 b+15\right) \Gamma_0 + 2 b\left((10-4 b) \Gamma_1 + b \Gamma_2\right)\right), \\
         G_{3} &= \frac{1}{2} \left(b \Gamma_1-(b-2) \Gamma_0\right), \\
      G_{4} &= \frac{1}{4} \left(\left(3 b^2-11 b+10\right) \Gamma_0+b \left((11-4 b) \Gamma_1+b \Gamma_2\right)\right),\\
      G_{5} &= \frac{1}{8} \left(\left(3 b^2-12 b+14\right) \Gamma_0+b \left(b \Gamma_2-4 (b-3) \Gamma_1\right)\right),
\end{align}
where $\Gamma_{n}(b) = I_n(b)\exp(-b)$ and $I_n$ is the modified Bessel function of the first kind. 

The equations which arise upon taking velocity space moments of the optimal electron distribution are, 
\begin{multline}
\label{eqn:1stelectronmoment}
    2\Lambda\Bigg((1 + \tau)\kappa_{1e} - \frac{\tau}{2(1 + \tau)}\int_{1/B_\mathrm{max}}^{1/B(l)} \frac{B(l) \, \mathrm{d}\lambda}{\sqrt{1 - \lambda B}}\Bigg\{ \overline{\alpha (\kappa_{1i} - \kappa_{1e})}  \\[5pt]
     - \frac{\Delta\tau}{2}\overline{\int_{1/B_\mathrm{max}}^{1/B(l)} \frac{B(l) \, \mathrm{d}\lambda}{\sqrt{1 - \lambda B}} (\overline{\kappa_{1i}}- \overline{\kappa_{1e}})} \Bigg\} \Bigg) =\frac{\tau i}{2}\int_{1/B_\mathrm{max}}^{1/B(l)} \frac{B(l) \, \mathrm{d}\lambda}{\sqrt{1 - \lambda B}}\Bigg(\frac{3}{2}\omega_{*e}\eta_e (\overline{\kappa_{1e}} - \overline{\kappa_{1i}}) \\[5pt]
     - \omega_{*e}(1 - 3\eta_e/2)\overline{\kappa_{1i}}
    - \omega_{*e}\eta_e \overline{\kappa_{2e}} + \omega_{*i}(1-3\eta_i/2)\overline{\kappa_{1i}} + \omega_{*i}\eta_i \overline{\kappa_{2i}} \\[5pt]
    -\Delta\bigg\{\overline{\hat{\omega}_{di}\kappa_{3i}} - \tilde{\omega}_{de}\overline{\kappa_{3e}} -\frac{3}{2}\overline{\hat{\omega}_{de}(1 - \lambda B/2)}(\overline{\kappa_{1i}} - \overline{\kappa_{1e}}) \bigg\}\Bigg),
\end{multline}

\begin{multline}
\label{eqn:2ndelectronmoment}
 2\Lambda\Bigg((1 + \tau)\kappa_{2e} - \frac{3\tau}{4(1 + \tau)}\int_{1/B_\mathrm{max}}^{1/B(l)} \frac{B(l) \, \mathrm{d}\lambda}{\sqrt{1 - \lambda B}}\Bigg\{ \overline{\alpha (\kappa_{1i} - \kappa_{1e})} \\[5pt] 
    - \frac{\Delta\tau}{2}\overline{\int_{1/B_\mathrm{max}}^{1/B(l)} \frac{B(l) \, \mathrm{d}\lambda}{\sqrt{1 - \lambda B}} (\overline{\kappa_{1i}}- \overline{\kappa_{1e}})} \Bigg\} \Bigg) =\frac{3\tau i}{4}\int_{1/B_\mathrm{max}}^{1/B(l)} \frac{B(l) \, \mathrm{d}\lambda}{\sqrt{1 - \lambda B}}\Bigg(\frac{5}{2}\omega_{*e}\eta_e (\overline{\kappa_{1e}} - \overline{\kappa_{1i}})\\[5pt]
    - \omega_{*e}(1 - 3\eta_e/2)\overline{\kappa_{1i}}
    - \omega_{*e}\eta_e \overline{\kappa_{2e}} + \omega_{*i}(1-3\eta_i/2)\overline{\kappa_{1i}} + \omega_{*i}\eta_i \overline{\kappa_{2i}} \\[5pt]
    -\Delta\bigg\{\overline{\hat{\omega}_{di}\kappa_{3i}} - \tilde{\omega}_{de}\overline{\kappa_{3e}} -\frac{5}{2}\overline{\hat{\omega}_{de}(1 - \lambda B/2)}(\overline{\kappa_{1i}} - \overline{\kappa_{1e}}) \bigg\}\Bigg),
\end{multline}
\begin{multline}
\label{eqn:3rdelectronmoment}
 2\Lambda\Bigg((1 + \tau)\kappa_{3e} - \frac{3\tau}{4(1 + \tau)}\int_{1/B_\mathrm{max}}^{1/B(l)} \frac{B(l) \, \mathrm{d}\lambda}{\sqrt{1 - \lambda B}}\overline{f(l)(1 -\lambda B/2)}\Bigg\{ \overline{\alpha (\kappa_{1i} - \kappa_{1e})} \\[5pt] 
    - \frac{\Delta\tau}{2}\overline{\int_{1/B_\mathrm{max}}^{1/B(l)} \frac{B(l) \, \mathrm{d}\lambda}{\sqrt{1 - \lambda B}} (\overline{\kappa_{1i}}- \overline{\kappa_{1e}})} \Bigg\} \Bigg) =\frac{3\tau i}{4}\int_{1/B_\mathrm{max}}^{1/B(l)} \frac{B(l) \, \mathrm{d}\lambda}{\sqrt{1 - \lambda B}} \overline{f(l)(1 -\lambda B/2)} \Bigg(\\[5pt]
   \frac{5}{2}\omega_{*e}\eta_e (\overline{\kappa_{1e}} - \overline{\kappa_{1i}}) - \omega_{*e}(1 - 3\eta_e/2)\overline{\kappa_{1i}}
    - \omega_{*e}\eta_e \overline{\kappa_{2e}} + \omega_{*i}(1-3\eta_i/2)\overline{\kappa_{1i}} + \omega_{*i}\eta_i \overline{\kappa_{2i}} \\[5pt]
    -\Delta\bigg\{\overline{\hat{\omega}_{di}\kappa_{3i}} - \tilde{\omega}_{de}\overline{\kappa_{3e}} -\frac{5}{2}\overline{\hat{\omega}_{de}(1 - \lambda B/2)}(\overline{\kappa_{1i}} - \overline{\kappa_{1e}}) \bigg\}\Bigg).
\end{multline}

\section{Eigenvalue problem in the square magnetic well}
\label{appendix:square_well}
In the slow-ion-transit limit with a square magnetic trapping well, the $l$-dependence of the eigenvalue problem \eqref{eqn:kinetic_eigenvalue_problem}  is trivial. This makes the system of  moment equations purely algebraic, allowing us to adopt a `super-moment' representation akin to that of Part II.  We define,
\begin{equation}
    \sigma_a = e_a n\left(\sum_a \frac{n e_a^2}{T_a}\right)^{-1/2},
\end{equation}
\begin{equation}
    \tilde{\kappa}_1 = \sum_a \sigma_a \kappa_{1a},
\end{equation}
\begin{equation}
    \tilde{\kappa}_2 = \sum_a \sigma_a{\omega}_{*a}'\left((1 - 3 \eta_a /2))\kappa_{1a} + \eta_a \kappa_{2a}\right),
\end{equation}
\begin{equation}
    \tilde{\kappa}_3 = \sum_a \sigma_a {\omega}_{da}' \kappa_{3a},
\end{equation}
where we have introduced the normalisation $\omega_{*a}' = \omega_{*a}/\omega_{*i}$ and $\omega_{da}' = \hat{\omega}_{da}/\omega_{*i}$. We may also define the following velocity-space-dependent factors,
\begin{align}
\psi_{1a} = J_{0a} && \psi_{2a} = \frac{v^2}{v_{Ta}^2}J_{0a}
\end{align}
\begin{equation}
    \psi_{3a} = \left(\frac{v_\perp^2}{2 v_{Ta}^2} + \frac{v_\parallel^2}{v_{Ta}^2}\right)J_{0a},
\end{equation}
where we once again take $J_{0e} \approx 1$. If we adopt the compact notation,
\begin{align}
    \tilde{\kappa}_m = \sum_{n,b} c_{mn}^{(b)} \kappa_{nb}, && \mathcal{I}_{mn}^{(a)} = \frac{c_{mn}^{(a)}\sigma_a}{n T_a},
\end{align}
then \eqref{eqn:kinetic_eigenvalue_problem} for the species `$a$' can be written as,
\begin{multline}
\label{eqn:square_well_eigenvalue_problem}
    \frac{\Lambda}{\omega_{*i}}\left(g_a + \frac{\tilde{\alpha} F_{a0}}{n T_a}\left( -\sigma_a \psi_{1a}\tilde{\kappa}_1 \right) \right) = \frac{i}{2}\frac{F_{a0}}{n T_a}\bigg(\omega_{*a}'(1 - 3\eta_a/2)\sigma_a \psi_{1a} \tilde{\kappa}_1 + \omega_{*a}'\eta_a \sigma_a \psi_{2a}\tilde{\kappa}_1 \\
    - \sigma_a \psi_{1a}\tilde{\kappa}_2 - \Delta \left[\sigma_a \omega_{da}'\psi_{3a}\tilde{\kappa}_1 -\sigma_a \psi_{1a}\tilde{\kappa}_3\right] \bigg)
\end{multline}
Here, we have evaluated the integral,
\begin{equation}
\int_{1/B_\mathrm{max}}^{1/B(l)}\frac{B(l) \, \mathrm{d}\lambda}{\sqrt{1 - \lambda B}} = 2 \sqrt{ 1 - \frac{B(l)}{B_\mathrm{max}}} = 2 \varepsilon(l)
\end{equation}
and defined $\tilde{\alpha} = (\alpha - \Delta \tau\varepsilon(l))/(1 + \tau)$. We can take moments of equation \eqref{eqn:square_well_eigenvalue_problem} by multiplying by the constant $c_{mn}^{(a)}$  and the function $\psi_{ma}$, summing over the index $n$ and the species label, and integrating over velocity space to arrive at,
\begin{multline}
    \frac{\Lambda}{\omega_{*i}}\left( \tilde{\kappa}_m  + \tilde{\alpha}\sum_{a,n} \left\{\mathcal{I}_{mn}^{(a)} X_{1n}^{(a)}\tilde{\kappa}_1 \right\}    \right) = \sum_{a,n} \frac{i}{2} \bigg( \omega_{*a}'(1 - 3/2\eta_a) \mathcal{I}_{mn}^{(a)} X_{1n}^{(a)}\tilde{\kappa}_1 \\
    + \omega_{*a}'\eta_a \mathcal{I}_{mn}^{(a)}X_{2n}^{(a)}\tilde{\kappa_1} - \mathcal{I}_{mn}^{(a)}X_{1n}^{(a)}\tilde{\kappa}_2 - \Delta\left[ \omega_{da}' \mathcal{I}_{mn}^{(a)} X_{3n}^{(a)}\tilde{\kappa}_1 - \mathcal{I}_{mn}^{(a)}X_{1n}^{(a)}\tilde{\kappa}_3\right]\bigg),
\end{multline}
where the factors $X_{mn}^{(a)}$ are integrals over velocity space defined by,
\begin{equation}
    X_{mn}^{(a)} = \frac{1}{n}\int F_{0a} \psi_{ma}\psi_{na} J_{0a}\, \mathrm{d}^3 v,
\end{equation}
where, for the electron species, the velocity space integral is carried out in the trapped region of velocity space, defined by $1/B_{\mathrm{max}} \leq \lambda \leq  1/B_{\mathrm{min}}$.

Written in this form, the system reduces to a simple $3\times 3$ generalised eigenvalue problem which may be solved for the eigenvalue $\Lambda$. The $X_{nm}^{(a)}$'s are given by,

\begin{align}
    & X_{11}^{(e)} = \varepsilon, &&  X_{11}^{(i)} = G_0(b),\\ 
        &X_{12}^{(e)} = \frac{3}{2}\varepsilon,  && X_{12}^{(i)} = G_1(b), \\
        & X_{13}^{(e)} = \frac{3}{4}\left(\varepsilon + \frac{1}{3}\varepsilon^3\right), && X_{13}^{i} = G_3(b),\\
        & X_{22}^{(e)} = \frac{15}{4}\varepsilon,  && X_{22}^{(i)} = G_2(b), \\
       & X_{32}^{(e)} = \frac{15}{8}\left(\varepsilon + \frac{1}{3}\varepsilon^3\right), && X_{32}^{(i)} = G_4(b),\\
        & X_{33}^{(e)} =  \frac{15}{8}\left(\frac{\varepsilon}{2} + \frac{1}{3}\varepsilon^3 + \frac{1}{10}\varepsilon^5\right), && X_{33}^{(i)} = G_5(b),
\end{align}
Where the functions $G_n(b)$ are given in Appendix \ref{appendix:Moment_forms}.
In this simple limit, the eigenvalue problem can be solved analytically to yield,
\begin{multline}
        {\Lambda_{\mathrm{SW}}^2}= \big(C_{ee}^{**}\omega_{*e}^2 + C_{ie}^{**}\omega_{*e}\omega_{*i} + C_{ii}^{**}\omega_{*i}^2 + C_{ii}^{* d}\omega_{*i}\omega_{di} +C_{ii}^{d d}\omega_{di}^2 +  C_{ee}^{* d }\omega_{*e}\omega_{de} + C_{ee}^{d d}\omega_{de}^2 \\
+ C_{ie}^{* d}\omega_{*i}\omega_{de} + C_{ie}^{d*}\omega_{*e}\omega_{di} \big)\bigg/\left(16(1 + \tau)(1 + \tau - \tilde{\alpha}\left(X_{11}^{(i)}+ \tau X_{11}^{(e)}\right) \right), 
\label{eqn:square_well_solution}
\end{multline}
where we define the following factors,
\begin{multline}
    C_{ee}^{**} = \tau X_{11}^{(e)}((2 -3\eta_e)^2) X_{11}^{(i)} + 4 \tau \eta_e^2 X_{22}^{(e)}) \\ +4 \tau \eta_e (X_{11}^{(i)}((2-3\eta_e) X_{12}^{(e)} +\eta_e X_{22}^{(e)}) - \tau \eta_e {X_{12}^{(e)}}^2)
\end{multline}
\begin{equation}
    C_{ie}^{**} = -2\tau((3\eta_e -2)X_{11}^{(e)} -2\eta_eX_{12}^{(e)})((3\eta_i-2)X_{11}^{(i)} -2\eta_iX_{12}^{(i)})
\end{equation}
\begin{multline}
    C_{ii}^{**} = \tau X_{11}^{(e)}((2-3\eta_i)^2X_{11}^{(i)} + 4\eta_i((2 - 3\eta_i)X_{12}^{(i)} + \eta_i X_{22}^{(i)})) \\
    -4\eta_i^2({X_{12}^{(i)}}^2 - X_{11}^{(i)} X_{22}^{(i)})
\end{multline}
\begin{equation}
    C^{dd}_{ee} = 4 \Delta^2 \tau (X_{33}^{(e)}(X_{11}^{(i)} +\tau X_{11}^{(e)}) - \tau {X_{13}^{(e)}}^2)
\end{equation}
\begin{equation}
     C^{dd}_{ii} = 4 \Delta^2 (X_{33}^{(i)}(X_{11}^{(i)} +\tau X_{11}^{(e)}) - {X_{13}^{(i)}}^2)
\end{equation}
\begin{equation}
    C^{dd}_{ie} = -8 \tau \Delta^2 X_{13}^{(i)}X_{13}^{(e)}
\end{equation}
\begin{equation}
   C^{*d}_{ee} = 4 \Delta \tau (X_{11}^{(i)}((3\eta_e- 2) X_{13}^{(e)}-2\eta_e X_{23}^{(e)})+ 2\tau\eta_e(X_{12}^{(e)} X_{13}^{(e)} -X_{11}^{(e)} X_{23}^{(e)}))
\end{equation}
\begin{equation}
    C^{d *}_{i e} = 4 \Delta \tau (X_{13}^{(i)}(X_{11}^{(e)}(2-3\eta_e) + 2 X_{12}^{(e)} \eta_e))
\end{equation}

\begin{equation}
    C^{*d}_{ie} = 4 \Delta \tau ((X_{13}^{(e)}((X_{11}^{(i)}(2-3\eta_i)+2 (X_{12}^{(i)}\eta_i))
\end{equation}

\begin{equation}
    C^{*d}_{ii} = 4 \Delta (2 \eta_i(X_{12}^{(i)} X_{13}^{(i)} - X_{11}^{(i)} X_{23}^{(i)}) + \tau X_{11}^{(e)}((3\eta_i -2) X_{13}^{(i)} - 2 X_{23}^{(i)} \eta_i))
\end{equation}

\section{Linear Theory}
\label{Appendix:linear_theory}
To make a comparison between the upper bounds derived here in the bounce-averaged electron response limit, we must solve the integral equation,
\begin{equation}
    (1 + \tau - R_i(\xi_i, l))\phi = \tau \int_{1/B_\mathrm{max}}^{1/B(l)} \frac{B(l) \, \mathrm{d}\lambda}{\sqrt{1 - \lambda B} }R_e(\lambda, \xi_e) \overline{\phi}
\end{equation}
where $\xi_i = \omega/\hat{\omega}_{di}(l)$,  $\xi_e = \omega/(\overline{\hat{\omega}_{de}(1 -\lambda B/2)})$, and the ion response $R_i$ is given by the standard expression of \citet{biglariToroidalIonPressure1989},

\begin{equation}
    R_i(\xi_i, l) = Y(\xi_i)^2 + \kappa_{di} \left\{\left[\frac{\eta_i -1}{\xi_i} - 2\eta_i  \right]Y(\xi_i)^2 + 2\eta_i Y^2   \right\}
\end{equation}
where $\kappa_{di} = \omega_{*i}/\hat{\omega}_{di}(l)$ and $Y(\xi_i) = -\sqrt{\xi_i}Z(\sqrt{\xi_i})$ with $Z$ being the plasma dispersion function. The electron response $R_e$ is given by,
\begin{equation}
    R_e(\xi_e, \lambda) = (\kappa_{de} -\xi_e)\left\{1 + \frac{1}{2}\sqrt{\xi_e}\left[ Z(\sqrt{\xi_e}) - Z(\sqrt{\xi_i})\right]\right\},
\end{equation}
where $\kappa_{de} = \omega_{*e}/(\overline{\hat{\omega}_{de}(1 -\lambda B/2)})$ and we have taken $\eta_e = 0$ for simplicity.
For this work, we only consider the solution to this equation in a square magnetic trapping well with constant curvature. In this case, $\phi = \overline{\phi}$  and $\hat{\omega}_{da}(l) = \mathrm{const.}$ The integral over $\lambda$  and the roots of the dispersion relation can both be computed numerically.

\bibliographystyle{jpp}


\bibliography{bibliography}

\end{document}